\documentclass[sigconf]{acmart}

\usepackage{amsmath,amssymb,amsfonts}
\usepackage{algpseudocode}
\usepackage{graphicx}
\usepackage{array}
\usepackage{textcomp}
\usepackage{color, colortbl}
\usepackage{multirow}
\usepackage{subcaption}
\usepackage{listing}
\usepackage{url}
\usepackage[override]{cmtt}
\usepackage{listings}
\usepackage{booktabs}
\usepackage{enumitem}
\usepackage{balance}

\def\BibTeX{{\rm B\kern-.05em{\sc i\kern-.025em b}\kern-.08em
    T\kern-.1667em\lower.7ex\hbox{E}\kern-.125emX}}

\definecolor{orange}{RGB}{255,127,0}

\newcommand{\prog}[1]{\textit{#1}}
\begin{document}

\copyrightyear{2018} 
\acmYear{2018} 
\setcopyright{acmlicensed}
\acmConference[CCS '18]{2018 ACM SIGSAC Conference on Computer and Communications Security}{October 15--19, 2018}{Toronto, ON, Canada}
\acmBooktitle{2018 ACM SIGSAC Conference on Computer and Communications Security (CCS '18), October 15--19, 2018, Toronto, ON, Canada}
\acmPrice{15.00}
\acmDOI{10.1145/3243734.3243804}
\acmISBN{978-1-4503-5693-0/18/10}

\fancyhead{}

\title{Evaluating Fuzz Testing}

\author{George Klees, Andrew Ruef, Benji Cooper}
\affiliation{\institution{University of Maryland}}



\author{Shiyi Wei}
\affiliation{\institution{University of Texas at Dallas}}

\author{Michael Hicks}
\affiliation{\institution{University of Maryland}}

\begin{abstract}
  Fuzz testing has enjoyed great success at discovering security
  critical bugs in real software. Recently, researchers have devoted
  significant effort to devising new fuzzing techniques, strategies,
  and algorithms. Such new ideas are primarily evaluated
  experimentally so an important question is: What experimental setup is
  needed to produce trustworthy results? We surveyed the recent
  research literature and assessed the experimental evaluations
  carried out by 32 fuzzing papers. We found problems in every
  evaluation we considered. We then performed our own extensive
  experimental evaluation using an existing fuzzer. Our results showed
  that the general problems we found in existing experimental
  evaluations can indeed translate to actual wrong or misleading
  assessments. We conclude with some guidelines that we hope will help
  improve experimental evaluations of fuzz testing algorithms, making reported
  results more robust.
\end{abstract}

\ccsdesc[500]{Security and privacy}
\ccsdesc[500]{Security and privacy~Software and application security}

\keywords{fuzzing, evaluation, security}

\maketitle


\section{Introduction}

A \emph{fuzz tester} (or \emph{fuzzer}) is a tool that iteratively and
randomly generates inputs with which it tests a target
program. Despite appearing ``naive'' when compared to more
sophisticated tools involving SMT solvers, symbolic execution, and
static analysis, fuzzers are surprisingly effective. For
example, the popular fuzzer AFL has been used to find hundreds of bugs
in popular programs~\cite{afl}. Comparing AFL head-to-head with the
symbolic executor \textit{angr}, AFL found 76\% more bugs (68 vs. 16)
in the same corpus over a 24-hour
period~\cite{stephens2016driller}. The success of fuzzers has made
them a popular topic of research.

Why do we think fuzzers work?  While inspiration for new ideas may be drawn
from mathematical analysis, fuzzers are primarily evaluated
experimentally.  When a researcher develops a new fuzzer 
algorithm (call it $A$), they must empirically demonstrate that it provides
an advantage over the status quo. To do this, they must choose:
\begin{itemize}
\item a compelling \emph{baseline} fuzzer $B$ to compare against;
\item a sample of target programs---the \emph{benchmark suite};
\item a \emph{performance metric} to
  measure when $A$ and $B$ are run on the benchmark suite; ideally,
  this is the number of (possibly exploitable) bugs identified by crashing inputs;
\item a meaningful set of \emph{configuration parameters}, e.g.,
  the \emph{seed file} (or files) to start fuzzing with, and the
  \emph{timeout} (i.e., the duration) of a fuzzing run.
\end{itemize}
An evaluation should also account for the fundamentally random nature
of fuzzing: Each fuzzing run on a target program may produce different
results than the last due to the use of randomness. As such, an
evaluation should measure \emph{sufficiently many trials} to sample the overall
distribution that represents the fuzzer's performance, using a
\emph{statistical test} \cite{Ott:2006:ISM:1203556} to determine that $A$'s measured improvement
over $B$ is real, rather than due to chance.

Failure to perform one of these steps, or failing to follow
recommended practice when carrying it out, could lead to misleading or
incorrect conclusions. Such conclusions waste time for practitioners,
who might profit more from using alternative methods or
configurations. They also waste the time of researchers, 
who make overly strong assumptions based on an arbitrary tuning of 
evaluation parameters.

We examined 32 recently published papers on fuzz testing (see
Table~\ref{table:sum}) located by perusing top-conference proceedings
and other quality venues, and studied their experimental
evaluations. We found that no fuzz testing evaluation
carries out all of the above steps properly (though some get
close).
This is bad news in theory, and after carrying out more than
50000 CPU hours of experiments, we believe it is bad news in practice,
too. Using AFLFast~\cite{Bohme:2016:CGF:2976749.2978428} (as $A$) and
AFL (as baseline $B$), we carried out a variety of tests of their
performance. We chose AFLFast as it was a recent advance over the 
state of the art; its code was publicly available; and we were
confident in our ability to rerun the experiments described by the authors
in their own evaluation and expand these experiments by varying
parameters that the original experimenters did not.
This choice was also driven by the importance of AFL in the literature:
14 out of 32 papers we examined used AFL as a baseline in their evaluation.
We targeted three binutils programs (\prog{nm},
\prog{objdump}, and \prog{cxxfilt}) and two image processing programs (\prog{gif2png} and
  \prog{FFmpeg}) used in prior fuzzing
  evaluations~\cite{Woo:2013:SBM:2508859.2516736,
    rawat2017vuzzer, Cha:2015:PMF:2867539.2867674,
    Rebert:2014:OSS:2671225.2671280, DBLP:conf/cis/ZhangYFT17}.
We found that experiments that deviate 
from the above recipe could easily lead one to draw incorrect
conclusions, for these reasons:

\emph{Fuzzing performance under the same configuration can vary
  substantially from run to run}. Thus, comparing single runs, as
nearly $\frac{3}{5}$ of the
  examined papers seem to, does not give a full picture. For
  example, on \prog{nm}, one AFL run found just over 1200 crashing
  inputs while one AFLFast run found around 800. Yet, comparing
  the median of 30 runs tells a different story: 400 crashes for AFL
  and closer to 1250 for AFLFast. Comparing averages is still not enough,
  though: We found that in some cases, via a statistical test, that
  an apparent difference in performance was not statistically significant.

\emph{Fuzzing performance can vary over the course of a run}. This
means that short timeouts (of less than 5 or 6 hours, as used by 11 papers) 
may paint a misleading picture. For example, when using the empty seed, AFL
  found no crashes in \prog{gif2png} after 13 hours, while AFLFast had
  found nearly 40. But after 24 hours AFL had found 39 and AFLFast had
  found 52. When using a non-empty seed set, on \prog{nm} AFL
  outperformed AFLFast at 6 hours, with statistical
  significance, but after 24 hours the trend reversed. 

We similarly
  found \emph{substantial performance variations based on the seeds used};
  e.g., with an empty seed AFLFast found more than 1000 crashes in
  \prog{nm} but with a small non-empty seed it found only 24, which was
  statistically indistinguishable from the 23 found by AFL. And yet,
  most papers treated the choice of seed casually, apparently assuming
  that any seed would work equally well, without providing
  particulars.

Turning to measures of performance, \emph{14 out of
    32 papers we examined used code coverage to assess fuzzing
    effectiveness.} Covering more code intuitively correlates with finding more
  bugs~\cite{7081877,Gopinath:2014:CCS:2568225.2568278} and so would
  seem to be worth doing. But the correlation may be
  weak~\cite{Inozemtseva:2014:CSC:2568225.2568271}, so directly
  measuring the number of bugs found is preferred. \emph{Yet only
    about $\frac{1}{4}$ of papers used this 
    direct measure}. Most papers instead counted the number of
  \emph{crashing inputs} 
  found, and then applied a heuristic procedure in an attempt to
  de-duplicate inputs that trigger the same bug (retaining a
  ``unique'' input for that bug). The two most popular
  heuristics were AFL's coverage profile (used by 7 papers) and
  (fuzzy) stack hashes~\cite{Molnar:2009:DTG:1855768.1855773} (used by
  7 papers). Unfortunately, there is reason to believe these
  \emph{de-duplication heuristics are ineffective.} 

In an additional experiment we computed a portion of ground truth. We applied all
  patches to \prog{cxxfilt} from the version we fuzzed up until the
  present. We grouped together all inputs that a particular patch
  caused to now gracefully exit~\cite{Chen:2013:TCF:2491956.2462173},
  confirming that the patch represented a single conceptual bugfix.
  We found that all  
  57,142 crashing inputs deemed ``unique'' by coverage profiles were
  addressed by 9 distinct patches. This
  represents a dramatic overcounting of the number of
  bugs. Ultimately, while AFLFast found many more ``unique'' crashing
  inputs than AFL, it only had a slightly higher likelihood of finding
  more unique bugs in a given run. 

Stack
  hashes did better, but still over-counted bugs. Instead of the bug mapping
  to, say 500 AFL coverage-unique crashes in a given trial, it would
  map to about 46 stack hashes, on average. Stack hashes were
  also subject to false negatives: roughly 16\% of hashes for crashes
  from one bug were shared by crashes from another bug. In five cases,
  a distinct bug was found by only one crash, and that crash had a
  non-unique hash, meaning that evidence of a distinct bug would have
  been dropped by ``de-duplication.'' 

  This experiment, the most substantial of its kind, suggests that
  reliance on heuristics for evaluating performance is unwise. A
  better approach is to measure against ground truth directly by
  assessing fuzzers against known bugs, as we did above, or by
  using a synthetic suite such as CGC~\cite{cgc} or
  LAVA~\cite{DBLP:conf/sp/Dolan-GavittHKL16}, as done by 6 papers we
  examined. (8 other papers considered ground truth in part, but often
  as ``case studies'' alongside general claims made using inputs
  de-duplicated by stack hashes or coverage profiles.)

  Overall, \emph{fuzzing performance may vary with the target
    program}, so it is important to evaluate on a diverse,
  representative benchmark suite. In our experiments, we found that AFLFast
  performed generally better than AFL on \prog{binutils} programs
  (basically matching its originally published result, when using
  an empty seed), but
  did not provide a statistically significant advantage on the image
  processing programs. Had these programs been included in its
  evaluation, readers might have drawn more nuanced conclusions about
  its advantages. In general, \emph{few papers use a common, diverse
    benchmark suite}; about 6 used CGC or LAVA-M,
    and 2 discussed the methodology
    in collecting real-world programs, while the rest used
  a few handpicked programs, with little overlap in these choices
  among papers.
  The median number of real-world programs used in the evaluation was 7,
  and the most commonly used programs (\prog{binutils}) were shared by
  only four papers (and no overlap when versions are considered).
  As a result, individual
  evaluations may present misleading conclusions internally, and results are hard
  to compare across papers. 

Our study (outlined in Section~\ref{sec:overview}) suggests that meaningful scientific progress on fuzzing 
requires that claims of algorithmic
improvements be supported by more solid evidence. 
\emph{Every evaluation in the 32 papers we looked at lacks some
  important aspect in this regard}. In this paper we propose some clear
guidelines to which future papers' evaluations should adhere. In
particular, researchers should perform multiple trials and use
statistical tests (Section~\ref{sec:sta-sig});
they should evaluate different seeds (Section~\ref{sec:seed}), and
should consider longer ($\geq 24$ hour vs. $5$ hour)
timeouts (Section~\ref{sec:timeout}); and they should evaluate bug-finding
performance using ground truth rather than heuristics such as ``unique crashes''
(Section~\ref{sec:perf}). Finally, we argue for the establishment
and adoption of a good fuzzing benchmark, and sketch what it might
look like. The practice of hand selecting
a few particular targets, and varying them from paper to paper, is
problematic (Section~\ref{sec:bm}). A well-designed and agreed-upon
benchmark would address this problem. We also identify other
problems that our results suggest are worth studying, including the
establishment of better 
de-duplication heuristics (a topic of recent interest~\cite{vantonder18sbc,pham17bucketing}), and
the use of algorithmic ideas from related areas, such as SAT solving. 



\section{Background}
\label{background}

There are many different dynamic analyses that can be described as
``fuzzing.''  A unifying feature of fuzzers is that they operate on,
and produce, concrete inputs. Otherwise, fuzzers might be instantiated
with many different design choices and many different parameter
settings. In this section, we outline the basics of how fuzzers work, and then touch
on the advances of 32 recently published papers which form
the core of our study on fuzzing evaluations.

\subsection{Fuzzing Procedure}

\begin{figure}
\underline{Core fuzzing algorithm:}
\begin{algorithmic}
 \State corpus $\gets$ initSeedCorpus()
 \State queue $\gets \emptyset$ 
 \State observations $\gets \emptyset$
 \While { $\neg$isDone(observations,queue) }
   \State candidate $\gets$ choose(queue, observations) 
   \State mutated $\gets$ mutate(candidate,observations)
   \State observation $\gets$ eval(mutated)
   \If{ isInteresting(observation,observations) }
     \State queue $\gets$ queue $\cup$ mutated
     \State observations $\gets$ observations $\cup$ observation
   \EndIf
 \EndWhile
\end{algorithmic}

parameterized by functions:

\begin{itemize}
 \item \textbf{initSeedCorpus}: Initialize a new seed corpus.
 \item \textbf{isDone}: Determine if the fuzzing should stop or 
       not based on progress toward a goal, or a timeout.
 \item \textbf{choose}: Choose at least one candidate seed from the 
       queue for mutation.
 \item \textbf{mutate}: From at least one seed and any observations made 
       about the program so far, produce a new candidate seed.
 \item \textbf{eval}: Evaluate a seed on the program to produce an 
       observation.
 \item \textbf{isInteresting}: Determine if the observations produced
       from an evaluation on a mutated seed indicate that the input
       should be preserved or not. 
\end{itemize}
\caption{Fuzzing, in a nutshell}
\label{fig:fuzz-alg}
\end{figure}

Most modern fuzzers follow the procedure outlined in
Figure~\ref{fig:fuzz-alg}. The process begins by choosing a corpus of
``seed'' inputs with which to test the target program. 
The fuzzer then repeatedly mutates these inputs and evaluates 
the program under test. If the
result produces ``interesting'' behavior, the fuzzer keeps the mutated input for 
future use and records what was observed. Eventually the fuzzer stops, either due to reaching
a particular goal (e.g., finding a certain sort of bug) or reaching a
timeout. 

Different fuzzers record different observations when running the program under
test. In a ``black box'' fuzzer, a single observation is made: whether the
program crashed. 
In ``gray box'' fuzzing, observations also consist of intermediate information about
the execution, for example, the branches taken during execution as
determined by pairs of basic block identifiers
executed directly in sequence. ``White box''
fuzzers can make observations and modifications by exploiting the
semantics of application
source (or binary) code, possibly involving sophisticated reasoning. Gathering additional 
observations adds overhead. Different
fuzzers make different choices, hoping to trade higher overhead for
better bug-finding effectiveness.

Usually, the ultimate goal of a fuzzer is to generate an input
that causes the program to crash. In some fuzzer configurations, \emph{isDone}
checks the queue to see if there have been any crashes, and if there have been, 
it breaks the loop. Other fuzzer configurations seek to collect as many 
different crashes as they can, and so will not stop after the first
crash. For example, by default, \texttt{libfuzzer}~\cite{libfuzzer} will stop when it discovers 
a crash, while AFL will continue and attempt to discover different crashes.
Other types of observations
are also desirable, such as longer running times that could indicate
the presence of algorithmic complexity vulnerabilities~\cite{petsios2017slowfuzz}. 
In any of these cases, the output from the 
fuzzer is some concrete input(s) and configurations that can be used from 
outside of the fuzzer to reproduce the observation. This allows software developers
to confirm, reproduce, and debug issues.



\subsection{Recent Advances in Fuzzing}
\label{sec:advance}

The effectiveness of fuzz testing has made it an active area of
research. Performing a literature search we found 32 papers published
between 2012 and 2018 that propose and study improvements to various
parts of the core fuzzing algorithm; 25 out of 32 papers we examined
were published since 2016. To find these papers, we started from 10
high-impact fuzzing papers published in top security venues. Then we
chased citations to and from these papers. As a sanity check, we also
did a keyword search of titles and abstracts of the papers published
since 2012. Finally, we judged the relevance based on target domain
and proposed advance, filtering papers that did not fit. 

Table \ref{table:sum} lists
these papers in chronological order. Here we briefly summarize the
topics of these papers, organized by the part of the fuzzing procedure
they most prominently aim to improve. Ultimately, our interest is in
how these papers \emph{evaluate} their claimed improvements, as
discussed more in the next section. 

{\bf initSeedCorpus}.
Skyfire \cite{DBLP:conf/sp/WangCWL17} and Orthrus \cite{DBLP:conf/raid/ShastryLFTYRSSF17}
propose to improve the initial seed selection by
running an up-front analysis on the program to bootstrap
information both for creating the corpus and assisting the mutators.
QuickFuzz \cite{grieco2016quickfuzz, Grieco:2017:QTF:3163934.3163968}
allows seed generation through the use of grammars that
specify the structure of valid, or interesting, inputs.
DIFUZE performs an up-front static analysis to identify the structure
of inputs to device drivers prior to fuzzing \cite{DBLP:conf/ccs/CorinaMSSHKV17}.


{\bf mutate}.
SYMFUZZ \cite{Cha:2015:PMF:2867539.2867674} uses a symbolic executor
to determine the number of bits of a seed to \emph{mutate}.
Several other works change \emph{mutate} to be aware of taint-level observations
about the program behavior, specifically mutating inputs that are 
used by the program \cite{rawat2017vuzzer, Cha:2012:UMB:2310656.2310692, li2017steelix,angora}. 
Where other fuzzers use pre-defined data mutation strategies like bit flipping 
or rand replacement, 
MutaGen uses fragments of the program under test that parse or manipulate the input as mutators
through dynamic slicing \cite{Kargen:2015:TPA:2786805.2786844}.
SDF uses properties of the seeds themselves to guide mutation \cite{DBLP:conf/iccst/LinLHL15}.
Sometimes, a grammar is used to guide mutation \cite{DBLP:conf/smartgridcomm/YooS16,DBLP:conf/ccs/HanC17}.
Chizpurfle's \cite{DBLP:conf/issre/IannilloNCN17} mutator exploits
knowledge of Java-level 
language constructs to assist in-process fuzzing of Android system services.

{\bf eval}.
Driller \cite{stephens2016driller} and MAYHEM \cite{Cha:2012:UMB:2310656.2310692}
observe that some conditional guards in the
program are difficult to satisfy via brute force guessing, and so (occasionally) invoke a symbolic 
executor during the \emph{eval} phase to get past them.
S2F also makes use of a symbolic executor during \emph{eval} \cite{DBLP:conf/cis/ZhangYFT17}.
Other work focuses on increasing the speed of \emph{eval} by making changes to
the operating system \cite{Xu:2017:DNO:3133956.3134046} or using different 
low level primitives to observe the effect of executions 
\cite{DBLP:conf/uss/SchumiloAGSH17,DBLP:conf/raid/HendersonYJH017,DBLP:conf/ccs/HanC17}.
T-Fuzz \cite{tfuzz} will transform the program to remove checks on the input that
prevent new code from being reached. MEDS \cite{medsalloc} performs finer grained 
run time analysis to detect errors during fuzzing. 

{\bf isInteresting}. While most papers focus on the crashes,
some work changes \emph{observation} to consider different classes of program
behavior as interesting, e.g., longer running time \cite{petsios2017slowfuzz}, or differential
behavior \cite{DBLP:conf/sp/PetsiosTSKJ17}.
Steelix \cite{li2017steelix} and Angora \cite{angora} instrument the program
so that finer grained information about
progress towards satisfying a condition is exposed through \emph{observation}.
Dowser and VUzzer \cite{dowser,rawat2017vuzzer} uses a static analysis to 
assign different rewards to program points based on either a likely-hood
estimation that traveling through that point will result in a vulnerability, 
or for reaching a deeper point in the CFG. 

{\bf choose}.
Several works select the next input candidate based on whether it
reaches particular areas of the program \cite{rawat2017vuzzer,lemieux2017fairfuzz,bohme2017directed, Bohme:2016:CGF:2976749.2978428}. Other work explores different algorithms for selecting 
candidate seeds \cite{Woo:2013:SBM:2508859.2516736,Rebert:2014:OSS:2671225.2671280}.

\begin{table*}[t]
\small
\centering
\begin{tabular}{|c|c|c|c|c|c|c|c|c|} 
\hline
{\bf paper} & {\bf benchmarks} & {\bf baseline} & {\bf trials} & {\bf variance} & {\bf crash} & {\bf coverage} & {\bf seed} & {\bf timeout}\\
\hline
MAYHEM\cite{Cha:2012:UMB:2310656.2310692} & R(29) & &  &  & G & ? & N & - \\
\hline
FuzzSim\cite{Woo:2013:SBM:2508859.2516736} & R(101) & B & 100 & C  & S &  & R/M & 10D\\
\hline
Dowser\cite{dowser} & R(7) & O & ? & & O & & N & 8H \\
\hline
COVERSET\cite{Rebert:2014:OSS:2671225.2671280} & R(10) & O & &  & S, G* & ? & R & 12H\\
\hline
SYMFUZZ\cite{Cha:2015:PMF:2867539.2867674} & R(8) & A, B, Z &  &  & S &  & M & 1H\\
\hline
MutaGen\cite{Kargen:2015:TPA:2786805.2786844} & R(8) & R, Z &  &  & S & L & V & 24H \\
\hline
SDF\cite{DBLP:conf/iccst/LinLHL15} & R(1) & Z, O &  &  & O &  & V & 5D \\
\hline
Driller\cite{stephens2016driller} & C(126) & A & &  & G & L, E & N & 24H \\
\hline
QuickFuzz-1\cite{grieco2016quickfuzz} & R(?) & & 10 &  & ? & & G & - \\
\hline
AFLFast\cite{Bohme:2016:CGF:2976749.2978428} & R(6) & A & 8 &  & C, G* & & E & 6H, 24H \\
\hline
SeededFuzz\cite{DBLP:conf/tase/WangSZ16} & R(5) & O & &  & M & O & G, R & 2H \\
\hline
\cite{DBLP:conf/smartgridcomm/YooS16} & R(2) & A, O & & & & L, E & V & 2H\\
\hline
AFLGo\cite{bohme2017directed} & R(?) & A, O & 20 &  & S & L & V/E & 8H, 24H\\
\hline
VUzzer\cite{rawat2017vuzzer} &C(63), L, R(10) & A & &  & G, S, O &  & N & 6H, 24H \\
\hline
SlowFuzz\cite{petsios2017slowfuzz} & R(10) & O & 100 &  & - &  & R/M/N & \\
\hline
Steelix\cite{li2017steelix} & C(17), L, R(5) & A, V, O & &  & C, G & L, E, M & N & 5H \\
\hline
Skyfire\cite{DBLP:conf/sp/WangCWL17} & R(4) & O & &  & ? & L, M & R, G & LONG\\
\hline
kAFL\cite{DBLP:conf/uss/SchumiloAGSH17} & R(3) & O & 5 &  & C, G* &  & V & 4D, 12D\\
\hline
DIFUZE\cite{DBLP:conf/ccs/CorinaMSSHKV17} & R(7) & O & &  & G* &  & G & 5H\\
\hline
Orthrus\cite{DBLP:conf/raid/ShastryLFTYRSSF17} & G(4), R(2) & A, L, O & 80 & C  & S, G* &  & V & $>$7D\\
\hline
Chizpurfle\cite{DBLP:conf/issre/IannilloNCN17} & R(1) & O & &  & G* &  & G & -\\
\hline
VDF\cite{DBLP:conf/raid/HendersonYJH017} & R(18) & & & & C & E & V & 30D\\
\hline
QuickFuzz-2\cite{Grieco:2017:QTF:3163934.3163968} & R(?) & O & 10 &  & G* &  & G, M & \\
\hline
IMF\cite{DBLP:conf/ccs/HanC17} & R(1) & O &  &  & G* & O & G & 24H\\
\hline
\cite{8227972} & S(?) & O & 5 &  & G &  & G & 24H\\
\hline
NEZHA\cite{DBLP:conf/sp/PetsiosTSKJ17} & R(6) & A, L, O & 100 &  & O &  & R & \\
\hline
\cite{Xu:2017:DNO:3133956.3134046} & G(10) & A, L & &  & &  & V & 5M\\
\hline
S2F\cite{DBLP:conf/cis/ZhangYFT17} & L, R(8) & A, O & &  & G & O & N & 5H, 24H\\
\hline
FairFuzz\cite{lemieux2017fairfuzz} & R(9) & A & 20 & C  & \- & E & V/M & 24H \\
\hline
Angora\cite{angora} & L, R(8) & A, V, O & 5 & & G, C & L, E & N & 5H \\
\hline
T-Fuzz\cite{tfuzz} & C(296), L, R(4) & A, O & 3 & & C, G* &  & N & 24H \\
\hline
MEDS\cite{medsalloc} & S(2), R(12)  & O & 10 & & C &  & N & 6H \\
\hline
\end{tabular}
\caption{\textmd{Summary of past fuzzing evaluation.
Blank cell means that the paper's evaluation did not mention this item;
- means it was not relevant; ? means the
element was mentioned but with insufficient detail to be clear about
it.
\textbf{Benchmarks}: R
    means real-world programs, C means CGC data-set, L means LAVA-M
    benchmark, S means programs with manually injected bugs, G means
    Google fuzzer test suite. \textbf{Baseline}:
    A means AFL, B means BFF \cite{bff}, L means libfuzzer
    \cite{libfuzzer}, R means Radamsa \cite{radamsa}, Z means Zzuf
    \cite{zzuf},
    V means VUzzer \cite{rawat2017vuzzer}
     O means other baseline used by no more than 1
    paper.
    \textbf{Trials}: number of trials.
    \textbf{Variance}: C means confidence intervals.   
    \textbf{Crash}: 
    S means stack hash used to group related crashes during triage, O means 
    other tools/methods used for triage, 
    C means coverage profile used to distinguish crashes,
    G means crashes triaged according to ground truth, G* means manual
    efforts partially obtained ground truth for triaging.
     \textbf{Coverage}: 
     L means
    line/instruction/basic-block coverage, M means method coverage, E
    means control-flow edge or branch coverage, O means other coverage
    information. \textbf{Seed}: R means randomly sampled seeds, M means
    manually constructed seeds, G means automatically generated seed,
    N means non-empty seed(s) but it was
    not clear if the seed corpus was valid,
    V means the paper assumes the existence
    of valid seed(s) but it was
    not clear how the seed corpus was obtained, E means empty seeds, /
    means different seeds were used in different programs, but only
    one kind of seeds in one program. \textbf{Timeout}: times reported
    in minutes (M), hours (H) and/or days (D).
    }} 
\label{table:sum}
\end{table*}

\section{Overview and Experimental Setup}
\label{sec:overview}

Our interest in this paper is assessing the existing research practice of
experimentally evaluating fuzz testing algorithms. As mentioned in the
introduction, evaluating a fuzz testing algorithm $A$
requires several steps: (a) choosing a baseline algorithm $B$ against
which to compare; (b) choosing a representative set of target programs
to test; (c) choosing how to measure $A$'s vs. $B$'s performance,
ideally as bugs found; (d) filling in algorithm parameters, such as
how seed files are chosen and how long the algorithm should run; and
(e) carrying out multiple runs for both $A$ and $B$ and statistically
comparing their performance.

Research papers on fuzz testing differ substantially
in how they carry out these steps. For each of the 32 papers
introduced in Section~\ref{sec:advance}, Table~\ref{table:sum}
indicates what \textbf{benchmark} programs were used for evaluation;
the \textbf{baseline} fuzzer used for comparison; the number of
\textbf{trials} carried out per configuration; whether
\textbf{variance} in performance was considered; how \textbf{crash}ing
inputs were mapped to bugs (if at all); whether code \textbf{coverage}
was measured to judge performance; how \textbf{seed} files were
chosen; and what \textbf{timeout} was used per trial (i.e., how long
the fuzzer was allowed to run). Explanations for each cell in the
table are given in the caption; a blank cell means that the paper's
evaluation did not mention this item.

For example, the AFLFast \cite{Bohme:2016:CGF:2976749.2978428} row
in Table \ref{table:sum} shows that the AFLFast's evaluation used 6 real-world
programs as benchmarks (column 2); used AFL as the baseline fuzzer (column 3);
ran each experiment 8 times (column 4) without reporting any variance (column 5);
measured and reported crashes, but also conducted manual triage to obtain ground
truth (column 6); did not measure code coverage (column 7);
used an empty file as the lone input seed (column 8);
and set 6 hours and 24 hours as timeouts for different experiments (column 9).

Which of these evaluations are ``good'' and which are not, in the
sense that they obtain evidence that supports the claimed technical
advance? In the following sections we assess evaluations both
theoretically and empirically, carrying out experiments that
demonstrate how poor choices can lead to misleading or incorrect
conclusions about an algorithm's fitness. In some cases, we believe it
is still an open question as to the ``best'' choice for an evaluation,
but in other cases it is clear that a particular approach
should be taken (or, at least, certain naive approaches should
\emph{not} be taken). Overall, we feel that \emph{every existing
  evaluation is lacking in some important way}.

We conclude this section with a description of the setup for our own
experiments. 

\paragraph*{Fuzzers} For our experiments we use AFL (with
standard configuration parameters) 2.43b as our baseline $B$, and
AFLFast~\cite{Bohme:2016:CGF:2976749.2978428} as our ``advanced''
algorithm $A$. We used the AFLFast version from July
2017 (cloned from Github) that was based on AFL version 2.43b. Note
that these are more recent versions than those used
in B\"{o}hme et al's original
paper~\cite{Bohme:2016:CGF:2976749.2978428}. Some, but not all, ideas
from the original AFLFast were incorporated into AFL by version
2.43b. This is not an issue for us since our goal is not to reproduce
AFLFast's results, but rather to use it as a representative
``advanced'' fuzzer for purposes of considering (in)validity of
approaches to empirically evaluating fuzzers. (We note, also, that AFL
served as the baseline for 14/32 papers we looked at, so using it in
our experiments speaks directly to those evaluations that used it.)
We chose it and AFL because they are open source, easy
to build, and easily comparable. We also occasionally consider a
configuration we call \emph{AFLNaive}, which is AFL with coverage
tracking turned off (using option \texttt{-n}), effectively turning
AFL into a black box fuzzer.

\paragraph*{Benchmark programs} We used the following benchmark programs
in our experiments: \prog{nm}, \prog{objdump}, \prog{cxxfilt} (all
from \prog{binutils-2.26}), \prog{gif2png}, and \prog{FFmpeg}.
All of these programs were obtained from
recent evaluations of fuzzing techniques. \prog{FFmpeg-n0.5.10} was
used in FuzzSim \cite{Woo:2013:SBM:2508859.2516736}. \prog{binutils-2.26} was
the subject of the AFLFast evaluation
\cite{Bohme:2016:CGF:2976749.2978428}, and only the three programs
listed above had discoverable bugs. \prog{gif2png-2.5.8} was tested by VUzzer
\cite{rawat2017vuzzer}.\footnote{Different versions of FFmpeg and gif2png were
assessed by other papers~\cite{Cha:2015:PMF:2867539.2867674,
    Rebert:2014:OSS:2671225.2671280,
    DBLP:conf/cis/ZhangYFT17}, and likewise for \prog{binutils}~\cite{lemieux2017fairfuzz,bohme2017directed,DBLP:conf/sp/PetsiosTSKJ17}.}
    We do not claim that this is a
complete benchmark suite; in fact, we think that a deriving a good
benchmark suite is an open problem. We simply use these programs to
demonstrate how testing on different targets might lead one to
draw different conclusions.

\paragraph*{Performance measure} For our experiments we measured the
number of ``unique'' crashes a fuzzer can induce over some period of
time, where uniqueness is determined by AFL's notion of coverage. In
particular, two crashing inputs are considered the same if they have
the same (edge) coverage profile. Though this measure is not uncommon,
it has its problems; Section \ref{sec:perf} discusses why, in detail.

\paragraph*{Platform and configuration} Our experiments were conducted
on three machines. Machines I and II are equipped with twelve 2.9GHz
Intel Xenon CPUs (each with 2 logical cores) and 48GB RAM running
Ubuntu 16.04. Machine III has twenty-four 2.4GHz CPUs and
110GB RAM running Red Hat Enterprise Linux Server 7.4. To account for possible
variations between these systems, each benchmark program was always
tested on the same machine, for all fuzzer combinations. Our testing script took
advantage of all the CPUs on the system to run as many trials in parallel as possible.
One testing subprocess was spawned per CPU and confined to it through CPU affinity.
 Every trial was
allowed to run for 24 hours, and we generally measured at least 30
trials per configuration. We also considered a variety of seed files,
including the empty file, randomly selected files of the right type,
and manually-generated (but well-formed) files.

\section{Statistically Sound Comparisons}
\label{sec:sta-sig}

All modern fuzzing algorithms fundamentally employ randomness when
performing testing, most notably when performing mutations, but
sometimes in other ways too. As such, it is not sufficient to simply
run fuzzer $A$ and baseline $B$ once each and compare their
performance. Rather, both $A$ and $B$ should be run for many trials,
and differences in performance between them should be judged. 

Perhaps surprisingly, Table~\ref{table:sum} shows that most (17 out of
32) fuzzing papers we considered make no mention of the number of
trials performed. Based on context clues, our interpretation is that
they each did one trial. One possible justification is that the
randomness ``evens out;'' i.e., if you run long enough, the random
choices will converge and the fuzzer will find the same number of
crashing inputs. It is clear from our experiments that this is not
true---fuzzing performance can vary dramatically from run to run.

\begin{figure*}[pt!]
 \centering
\begin{subfigure}{0.32\textwidth}
\centering
\includegraphics[width=\linewidth]{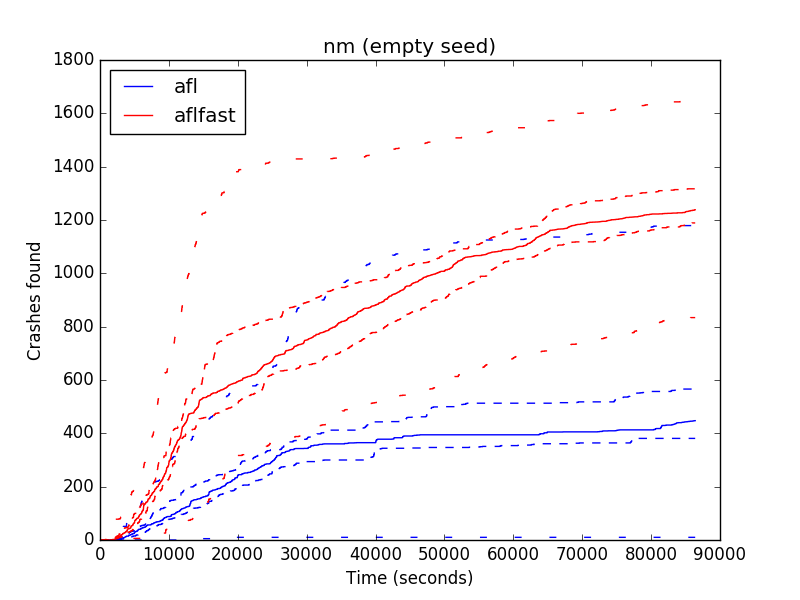}
\caption{nm: $p < 10^{-13}$}
\end{subfigure}
\begin{subfigure}{0.32\textwidth}
\centering
\includegraphics[width=\linewidth]{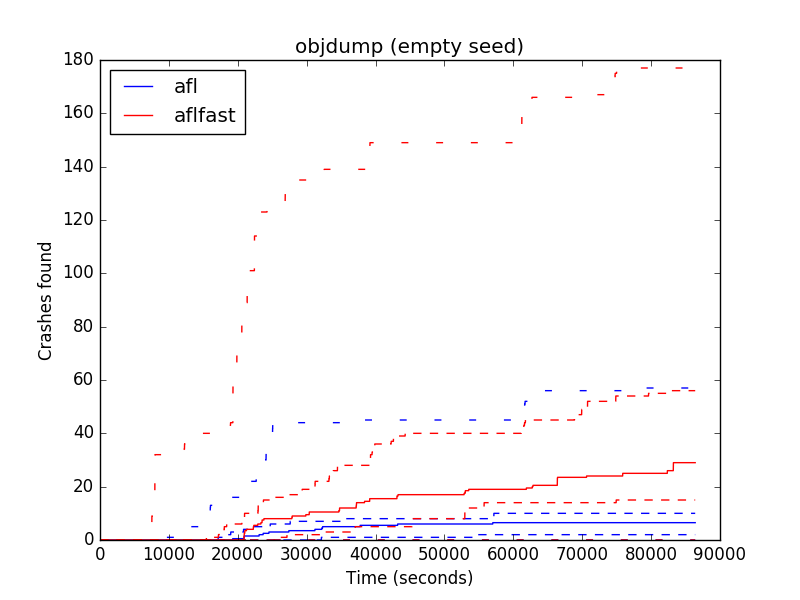}
\caption{objdump: $p < 0.001$}
\label{fig:objdump-empty}
\end{subfigure}
\begin{subfigure}{0.32\textwidth}
\centering
\includegraphics[width=\linewidth]{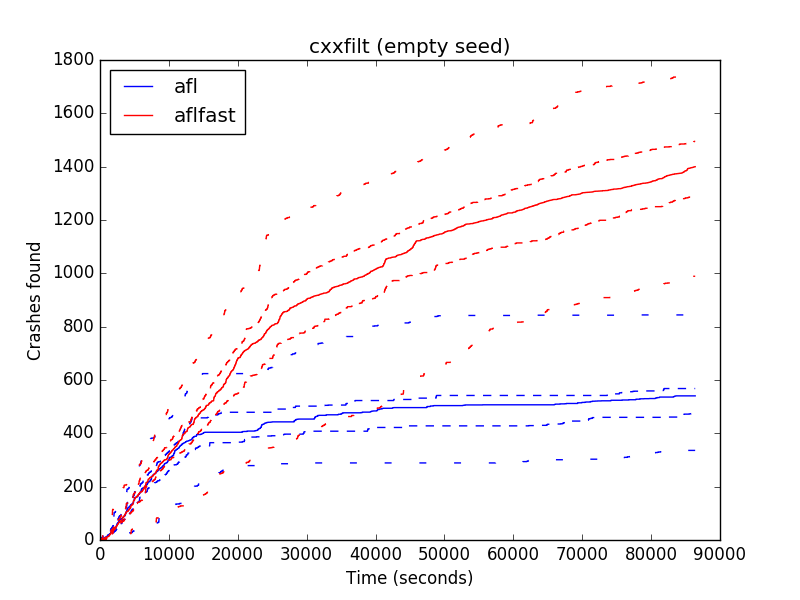}
\caption{cxxfilt: $p < 10^{-10}$}
\end{subfigure}
\begin{subfigure}{0.32\textwidth}
\centering
\includegraphics[width=\linewidth]{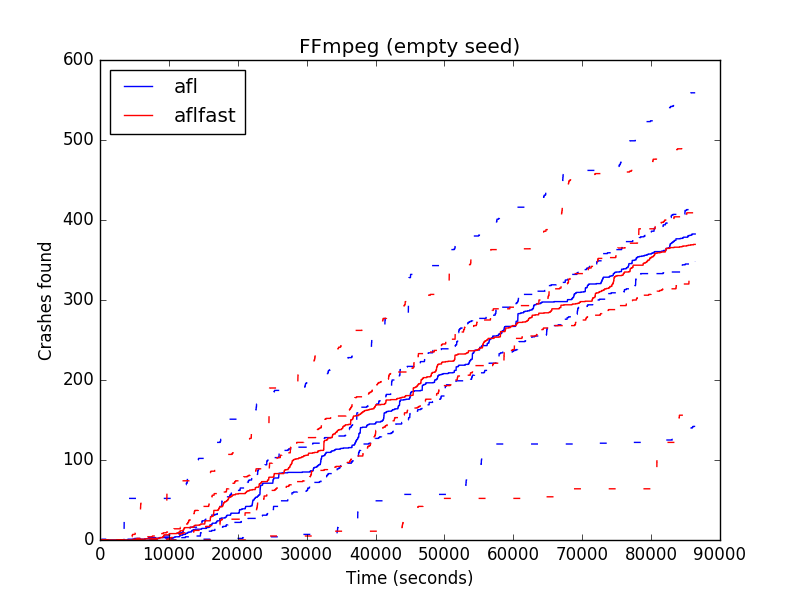}
\caption{FFmpeg: $p = 0.379$}
\label{fig:ffmpeg-empty}
\end{subfigure}
\begin{subfigure}{0.32\textwidth}
\centering
\includegraphics[width=\linewidth]{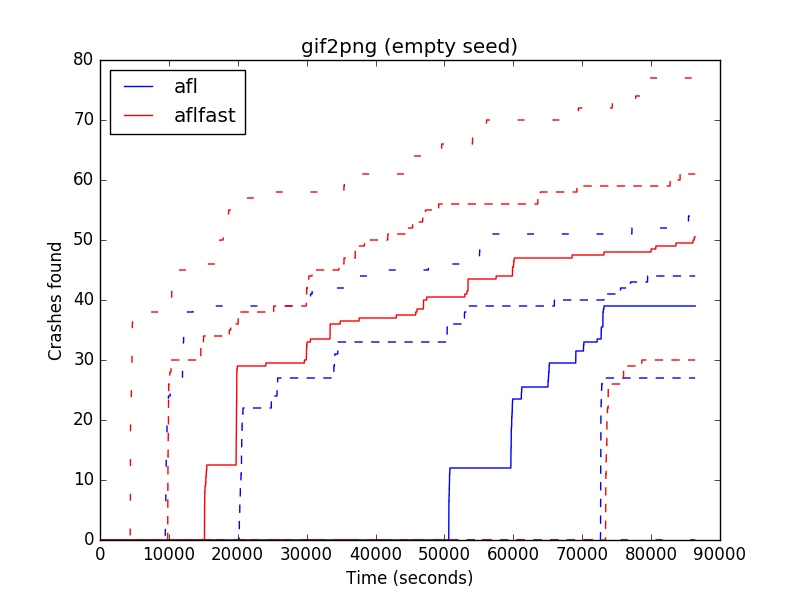}
\caption{gif2png: $p = 0.0676$}
\label{fig:gif2png-empty}
\end{subfigure}
\caption{Crashes found over time (empty seed). Solid line is median; dashed lines are confidence intervals, and max/min.}
\label{fig:result}
\end{figure*}

Consider the results presented in Figure~\ref{fig:result}, which
graphs the cumulative number of crashes (the Y axis) we found over time
(the X axis) by AFL (blue), and AFLFast (red), each starting with an
empty seed. In each plot, the solid line represents the median
result from 30 runs while the dashed lines represent the maximum and
minimum observed results, and the lower and
upper bounds of 95\% confidence intervals for a median~\cite{ci}. (The
outermost dashed lines are max/min, the inner ones are the CIs.) 

It should be clear from the highly varying performance on these plots that
considering only a single run could lead to the wrong conclusion. For
example, suppose the single run on \prog{FFmpeg} for AFL turned out to
be its maximum, topping out at 550 crashes, while the single run on
AFLFast turned out to be its minimum, topping out at only 150
crashes (Figure \ref{fig:ffmpeg-empty}). Just comparing these two results, we might believe that
AFLFast provides no advantage over AFL. Or we might
have observed AFLFast's maximum and AFL's minimum, and concluded the
opposite. 

Performing multiple trials and reporting averages is better, but not
considering variance is also problematic. In
Table~\ref{table:sum}, we can see that 12 out of the 15 papers that did
consider multiple trials did not characterize the performance
variance (they have a blank box in the \textbf{variance} column).
Instead, each of them compared the ``average'' performance (we assume: arithmetic
mean) of 
$A$ and $B$ when drawing conclusions, except for
Dowser \cite{dowser} that reported median, and 
two~\cite{grieco2016quickfuzz, 8227972} that did not mention how the
``average'' was calculated.  

The problem is that with a high enough variance, a difference in
averages may not be statistically significant. A solution 
is to use a \emph{statistical test}~\cite{Ott:2006:ISM:1203556}. Such a test
indicates the likelihood that a difference in performance is real,
rather than due to chance. Arcuri and
Briand~\cite{Arcuri:2011:PGU:1985793.1985795} suggest that for
randomized testing algorithms (like fuzzers), one should use the Mann
Whitney U-test to determine the stochastic ranking of $A$ and $B$,
i.e., whether the outcomes of the trials in $A$'s data sample are more
likely to be larger than outcomes in $B$'s. Mann Whitney is \emph{non
  parametric} in that it makes no assumption about the distribution of
a randomized algorithm's performance; by contrast, the standard
$t$-test assumes a normal distribution.

Returning to our experiments, we can see where simply comparing
averages may yield wrong conclusions. 
For example, for \prog{gif2png}, after 24 hours AFLFast finds a median
of 51 crashes while for AFL it is 39 (a difference of 12). But
performing the Mann Whitney test yields a $p$ value of greater than
$0.05$, suggesting the difference may not be statistically
significant, even on the evidence of thirty 24-hour trials. For \prog{FFmpeg},
AFLFast's median is 369.5 crashes while AFL's is 382.5, also a
difference of about 12 crashes, this time favoring AFL. Likewise, Mann
Whitney deems the difference insignificant. On the other hand, for
\prog{nm} the advantage of AFLFast over AFL is extremely unlikely to
occur by chance.

The three papers in Table~\ref{table:sum} with ``C'' in the
\textbf{variance} column come the closest to the best practice by
at least presenting confidence intervals along with averages. But even
here they stop short of statistically comparing the performance of
their approach against the baseline; they leave it to the reader to
visually judge this difference. This is helpful but not as conclusive
as a (easy to perform) statistical test. 

\begin{figure*}[th!]
 \centering
\begin{subfigure}{0.32\textwidth}
\centering
\includegraphics[width=\linewidth]{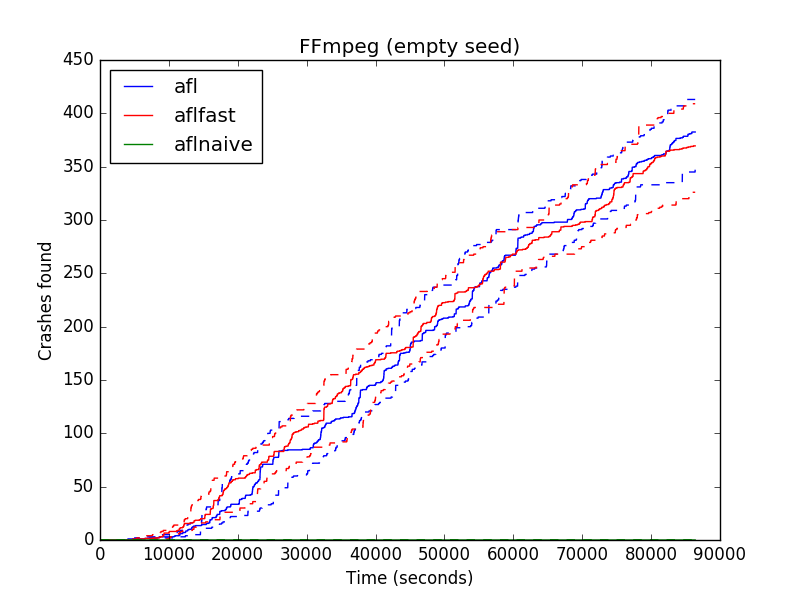}
\caption{empty seed\\$p_1 = 0.379$\\$p_2 < 10^{-15}$}
\end{subfigure}
\begin{subfigure}{0.32\textwidth}
\centering
\includegraphics[width=\linewidth]{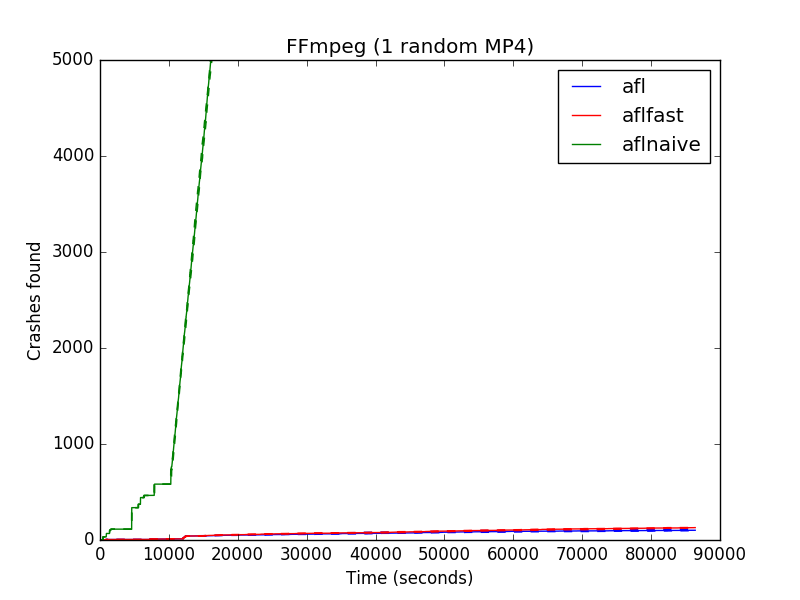}
\caption{1-made seed\\$p_1 = 0.048$\\$p_2 < 10^{-11}$}
\end{subfigure}
\begin{subfigure}{0.32\textwidth}
\centering
\includegraphics[width=\linewidth]{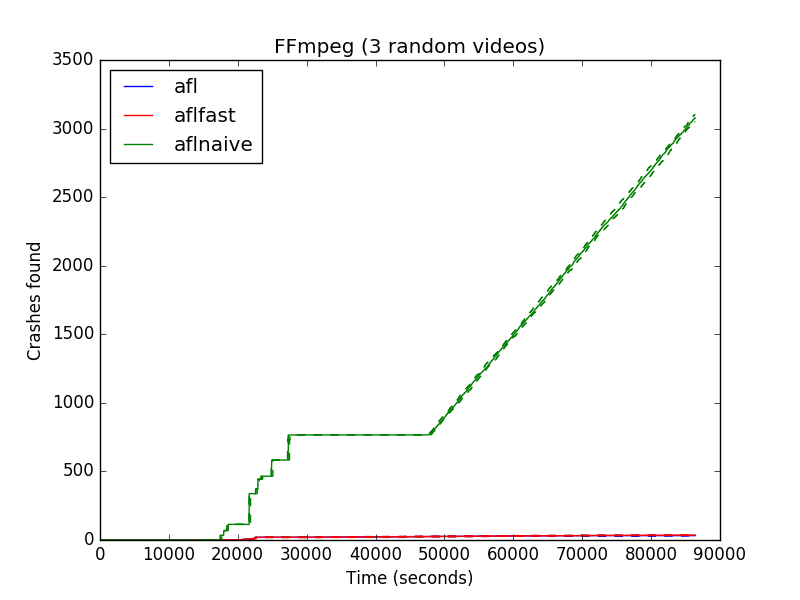}
\caption{3-made seeds\\$p_1 > 0.05$\\$p_2 < 10^{-10}$}
\end{subfigure}
\begin{subfigure}{0.32\textwidth}
\centering
\includegraphics[width=\linewidth]{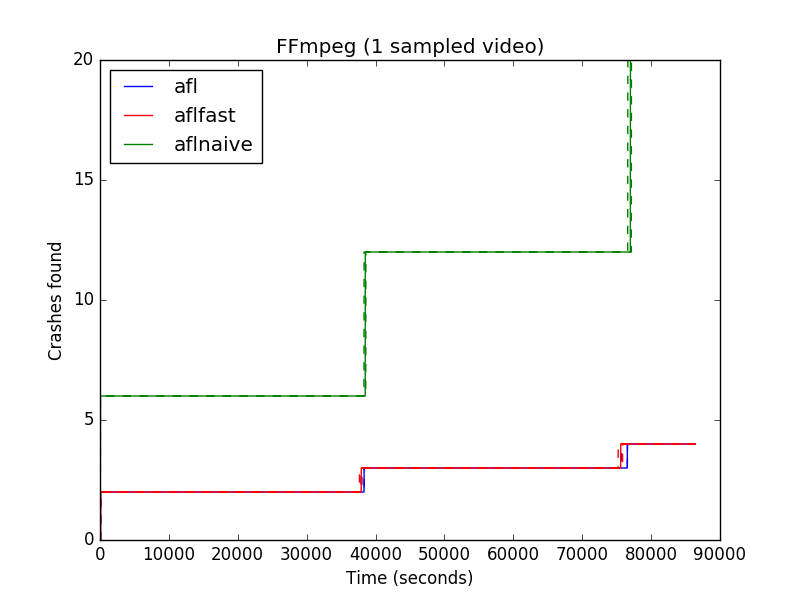}
\caption{1-sampled seeds\\$p_1 > 0.05$\\$p_2 < 10^{-5}$}
\end{subfigure}
\begin{subfigure}{0.32\textwidth}
\centering
\includegraphics[width=\linewidth]{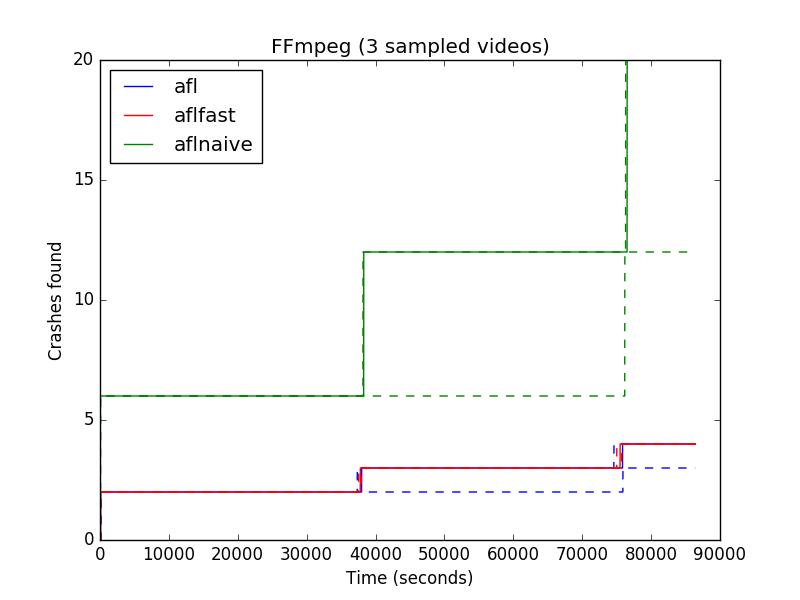}
\caption{3-sampled seeds\\$p_1 > 0.05$\\$p_2 < 10^{-5}$}
\end{subfigure}
\begin{subfigure}{0.32\textwidth}
\centering
\includegraphics[width=\linewidth]{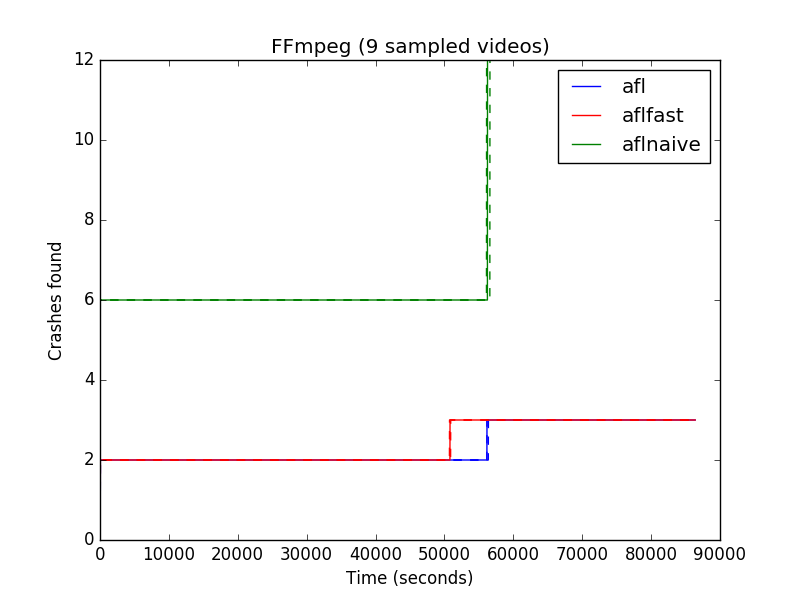}
\caption{9-sampled seeds\\$p_1 > 0.05$\\$p_2 < 10^{-6}$}
\end{subfigure}
\caption{\prog{FFmpeg} results with different seeds. Solid line is median result; dashed lines are confidence intervals. $p_1$ and $p_2$ are the p-values for the statistical tests of AFL vs. AFLFast and AFL vs. AFLNaive, respectively.}
\label{fig:ffmpeg}
\end{figure*}

\paragraph*{Discussion} While our recommendation to use statistical
tests should be uncontroversial, there can be further debate on the
best choice of test. In particular, two viable alternatives are the
\emph{permutation test}~\cite{drummond12perm} and
\emph{bootstrap}-based tests~\cite{calmettes12bootstrap}. These tests 
work by treating the measured data as a kind of stand-in for the
overall population, systematically comparing permutations and
re-samples of measured data to create rankings with confidence
intervals. Whether such methods are more or less appropriate than Mann
Whitney is unclear to us, so we follow Arcuri and Briand~\cite{Arcuri:2011:PGU:1985793.1985795}.

Determining that the median performance of fuzzer $A$ is greater than
fuzzer $B$ is paramount, but a related question concerns \emph{effect
  size}. Just because $A$ is likely to be better than $B$ doesn't tell
us \emph{how much} better it is. We have been implicitly answering
this question by looking at the difference of the measured
medians. Statistical methods could also be used to determine the
likelihood that this difference represents the true
difference. Arcuri and Briand
suggest Vargha and Delaney's $\hat{A}_{12}$
statistics~\cite{doi:10.3102/10769986025002101} (which employ elements
of the Mann Whitney calculation). Bootstrap methods can also be
employed here. 

%

\section{Seed Selection}
\label{sec:seed}

Recall from Figure~\ref{fig:fuzz-alg} that prior to iteratively
selecting and testing inputs, the fuzzer must choose an initial corpus
of \emph{seed} inputs.  Most (27 out of 32, per
Section~\ref{sec:advance}) recent papers focus on improving the main
fuzzing loop. As shown in column \textbf{seed} in Table
\ref{table:sum}, most papers (30/32) used a non-empty
seed corpus (entries with G, R, M, V, or N). 
A popular view is that a seed should be
well-formed (``valid'') and small---such seeds may drive the program to execute more of its
intended logic quickly, rather than cause it to terminate at its parsing/well-formedness
tests~\cite{rawat2017vuzzer,DBLP:conf/sp/WangCWL17,Rebert:2014:OSS:2671225.2671280,afl-readme}. And yet, many times the details of the particular
seeds used were not given. Entry 'V' appears 9
times, indicating a valid seed corpus was used, but providing no
details. Entry 'N' appears 10 times, indicating a non-empty seed, but
again with no details as to its content. Two
papers~\cite{Bohme:2016:CGF:2976749.2978428,bohme2017directed} opted
to use an empty seed (entry `E'). When
we asked them about it, they pointed out that using an empty seed is
an easy way to baseline a significant variable in the input
configuration. Other papers used manually or algorithmically constructed
seeds, or randomly sampled ones.

It may be that the details of the initial seed corpus are
unimportant; e.g., that no matter which seeds are used, algorithmic
improvements will be reflected. But it's also possible that there is a
strong and/or surprising interaction between seed format and algorithm
choice which could add nuance to the results~\cite{Mytkowicz:2009:PWD:1508244.1508275}. 
And indeed, this is what our results suggest.

We tested \prog{FFmpeg} with different seeds including the empty seed,
samples of existing video files (``sampled'' seeds)
and randomly-generated videos (``made'' seeds).
For the sampled seeds, videos were drawn from the \prog{FFmpeg}
samples website.\footnote{\url{http://samples.ffmpeg.org}} 
Four samples each were taken from the AVI, MP4, MPEG1, and MPEG2 sub-directories,
and then the files were filtered out to only include those less than 1 MiB,
AFL's maximum seed size, leaving {\it 9-sampled seeds} total.
This set was further pared down to the smallest of the video files to
produce {\it 3-sampled} and {\it 1-sampled} seeds.
For the made seeds, we generated video and GIF files
by creating 48 random video frames
with \texttt{videogen} (a tool included with \prog{FFmpeg}), 12 seconds of audio with \texttt{audiogen}
(also included), and stitching all of them together with \prog{FFmpeg}
into {\it 3-made} MP4, MPG, and AVI files, each at 4 fps.
The {\it 1-made} seed is the generated MP4 file.
We also tested \prog{nm}, \prog{objdump}, and \prog{cxxfilt} using the empty seed, and
a 1-made seed.
For \prog{nm} and \prog{objdump}, the 1-made seed was generated by
compiling a hello-world C program. The 1-made seed of \prog{cxxfilt} was generated
as a file with 16 random characters, chosen from
the set of letters (uppercase and lowercase), digits 0-9, and the underscore, which
is the standard alphabet of mangled C++ names.

Results with these different seed choices for \prog{FFmpeg} are shown
in Figure~\ref{fig:ffmpeg}. One clear trend is that for AFL and
AFLFast, the empty seed yields far more crashing inputs than any set of
valid, non-empty ones. On the other hand, for AFLNaive the trend is reversed.
Among the experiments with non-empty seeds, performance also
varies. For example, Figure~\ref{fig:ffmpeg}(b)
and Figure~\ref{fig:ffmpeg}(d) show very different performance with a
single, valid seed (constructed two different ways). The former finds
around 100 crashes for AFL and AFLFast after 24 hours, while the
latter finds less than 5.

\begin{table}[t]
\small
\centering
\begin{tabular}{|c|ll|ll|} 
\hline
 & \multicolumn{2}{c|}{{\bf empty}} & \multicolumn{2}{c|}{{\bf 1-made}} \\
\hline
FFmpeg, AFLNaive & 0 & ($<10^{-15}$) & 5000 & ($<10^{-11}$) \\
\hline
FFmpeg, AFL & 382.5 &  & 102 &  \\
\hline
FFmpeg, AFLFast & 369.5 & ($=0.379$) & 129 & ($<0.05$) \\
\hline
\hline
nm, AFL & 448 &  & 23 &  \\
\hline
nm, AFLFast & 1239 & ($<10^{-13}$) & 24 & ($=0.830$) \\
\hline\hline
objdump, AFL & 6.5 &  & 5&  \\
\hline
objdump, AFLFast & 29 & ($<10^{-3}$) & 6 & ($<10^{-2}$) \\
\hline \hline
cxxfilt, AFL & 540.5 &  & 572.5&  \\
\hline
cxxfilt, AFLFast & 1400 & ($<10^{-10}$) & 1364 & ($<10^{-10}$) \\ \hline
\end{tabular}
\caption{\textmd{Crashes found with different seeds. Median number of crashes at the 24-hour timeout.}}
\label{table:results}
\end{table}

The top part of Table~\ref{table:results} zooms in on the data from
Figure~\ref{fig:ffmpeg}(a) and (b) at the 24-hour mark. The first
column indicates the target program and fuzzer used; the second column
(``\textbf{empty}'') indicates the median number of crashes found when
using an empty seed; and the last column (``\textbf{1-made}'')
indicates the median number of crashes found when using a valid
seed. The parenthetical in the last two columns is the $p$-value for
the statistical test of whether the difference of AFLFast or AFLNaive
performance from AFL is real, or due to chance. 
For AFL and AFLFast, an empty seed produces hundreds of
crashing inputs, while for AFLNaive, it produces none. However, if we
use 1-made or 3-made seeds, AFLNaive found significantly more
crashes than AFL and AFLFast (5000 vs. 102/129).

The remainder of Table \ref{table:results} reproduces the results of
the AFLFast evaluation \cite{Bohme:2016:CGF:2976749.2978428} in the
\textbf{empty} column, but then reconsiders it with a valid seed in
the \textbf{1-made} column.  Similar to the conclusion made by the AFLFast
paper, AFLFast is superior to AFL in crash
finding ability when using the empty seed (with statistical
significance).  However, when using 1-made seeds, AFLFast is not quite as good: it no
longer outperforms AFL on \prog{nm}, and both AFL and AFLFast
generally find fewer crashes.

In sum, it is clear that a fuzzer's performance on the
same program can be very different depending on what seed is used.
Even valid, but different seeds can induce very different
behavior. Assuming that an evaluation is meant to show that fuzzer $A$
is superior to fuzzer $B$ \emph{in general}, our results suggest that
it is prudent to consider a variety of seeds when evaluating an
algorithm. Papers should be specific about how the seeds were
collected, and better still to make available the actual seeds
used. We also feel that the empty seed should be considered, despite
its use contravening conventional wisdom. In a sense, it is the most
general choice, since an empty file can serve as the input of
any file-processing program. If a fuzzer does well with the empty seed
across a variety of programs, perhaps it will also do well with the
empty seed on programs not yet tested. And it takes a significant
variable (i.e., which file to use as the seed) out of the
vast configuration space. 

\section{Timeouts}
\label{sec:timeout}

Another important question is how long to run
a fuzzer on a particular target. The last column of Table
\ref{table:sum} shows that prior experiments 
of fuzzers have set very different timeouts. These generally range from
1 hour to days and weeks.\footnote{\cite{Xu:2017:DNO:3133956.3134046}
 is an outlier that we do not count here: it uses 5-minute timeout
 because its evaluation focuses 
 on test generation rate instead of bug finding ability.}
 Common choices were 24 hours (10 papers) and
5 or 6 hours (7 papers). We observe that recent papers that used 
LAVA as the benchmark suite chose 5 hours
as the timeout, possibly because the same choice was made in 
the original LAVA paper~\cite{DBLP:conf/sp/Dolan-GavittHKL16}.
Six papers ran fuzzers for more than one day. 

\begin{figure}[pt!]
 \centering
\includegraphics[width=0.75\linewidth]{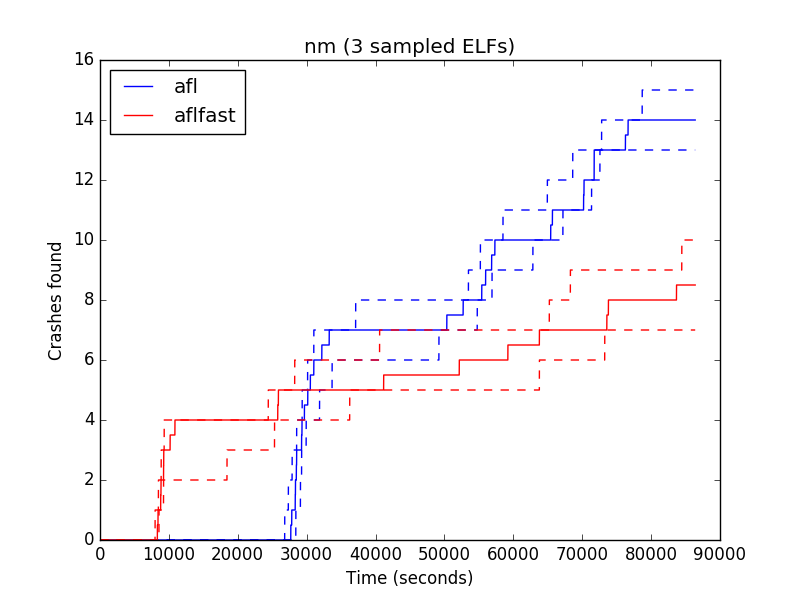}
\caption{\prog{nm} with three sampled seeds. At 6 hours: AFLFast is
  superior to AFL with $p < 10^{-13}$. At 24 hours: AFL is superior to
  AFLFast with $p = 0.000105$.}
\label{fig:timeout}
\end{figure}

Most papers we considered reported the timeout without justification. The
implication is that beyond a certain threshold, more running time is
not needed as the distinction between algorithms will be
clear. However, we found that relative performance between algorithms
can change over time, and that terminating an experiment too quickly
might yield an incomplete result. 
As an example, AFLFast's evaluation shows that AFL found
no bugs in \prog{objdump} after six hours \cite{Bohme:2016:CGF:2976749.2978428},
but running AFL longer seems
to tell a different story, as shown in Figure \ref{fig:objdump-empty}. After
six hours, both AFL and AFLFast start to find crashes at a reasonable
clip. Running AFL on \prog{gif2png} shows another interesting result
in Figure \ref{fig:gif2png-empty}. The median number of crashes found 
by AFL was 0 even after 13 hours, but with only 7 more hours, it found 40 crashes.
Because bugs often reside in certain parts of the program, fuzzing detects
the bugs only when these parts are eventually explored.
Figure \ref{fig:timeout} presents the results of AFL and AFLFast running
with three sampled seeds on \prog{nm}. 
After 6 hours none of the AFL runs found any bugs in \prog{nm},
while the median number of crashes found by AFLFast was 4; Mann Whitney
says that this difference is significant. But at 24 hours, the trend
is reversed: AFL has found 14 crashes and AFLFast only 8. Again, this
difference is significant.

What is a reasonable timeout to consider?
Shorter timeouts are convenient from a practical perspective, since
they require fewer overall hardware resources. Shorter times might be
more useful in certain real-world scenarios, e.g., as part of an
overnight run during the normal development process. On the other
hand, longer runs might illuminate more general performance
trends, as our experiments showed. Particular
algorithms might also be better with longer running times; e.g., they
could start slow but then accelerate their bug-finding ability as more
tests are generated. For example, Skyfire took several days before its better performance
(over AFL) became clear~\cite{DBLP:conf/sp/WangCWL17}. 

We believe that evaluations should include plots, as we have been (e.g., in Figure
\ref{fig:timeout}), that depict
performance over time. These runs should consider at least a 24 hour timeout;
performance for shorter times can easily be extracted from such
longer runs. 

\paragraph*{Discussion}

In addition to noting performance at particular times
(e.g., crash counts at 5, 8 and 24 hours), one could also report
\emph{area under curve} (AUC) as a less punctuated performance
measure. For example, a fuzzer that found one crash per second for
five seconds would have an AUC of 12.5 crash-seconds whereas a fuzzer
that found five crashes too, but all between seconds 4 and 5, would
have an AUC of 2.5 crash-seconds. These measures intuitively reflect
that finding crashes earlier and over time is preferred to
finding a late burst. On the other hand, this measure might prefer a steady
crash finder that peaks at 5 crashes to one that finds 10 at the last
gasp; aren't more crashes better? As such, AUC measures are not a
substitute for time-based performance plots.

\section{Performance Measures}
\label{sec:perf}

So far, we have focused on measuring fuzzer performance using
\emph{``unique'' crashes found}, which is to say, inputs that induce a
``unique'' crash (defined shortly). As crashes are symptoms of potentially serious
bugs, measuring a fuzzer according to the number of crashing inputs it
produces seems natural. But bugs and crashing inputs are not the same
thing: Many different inputs could trigger the same bug. For example,
a buffer overrun will likely cause a crash no matter what the input
data consists of so long as that data's size exceeds the buffer's
length. As such, simply counting crashing inputs as a measure of
performance could be misleading: fuzzer $A$ could find more crashes
than fuzzer $B$ but find the same or fewer actual bugs. 

As such, many papers employ some strategy to \emph{de-duplicate} (or
\emph{triage}) crashes, so as to map them to unique bugs. There are
two popular automated heuristics for doing this: using AFL's notion of
\emph{coverage profile}, and using \emph{stack hashes}. In
Table~\ref{table:sum}, these are marked `C' (7 papers) and `S' (7 papers) in the
\textbf{crash} column. 
There are four papers using other tools/methods for triage, marked `O'.
For example, VUzzer additionally
used a tool called \texttt{!Exploitable} to assess the exploitability
of a crash caused by a bug \cite{rawat2017vuzzer}. Crashes
that have a low likelihood of being turned into an attack could
be discounted by a user, so showing that a fuzzer finds
more dangerous bugs is advantageous. 
The de-duplication strategy used in our experiments corresponds
to `C'.

Unfortunately, as we show experimentally in this section, these
de-duplication heuristics are actually poor at clustering crashing
inputs according to their root cause. 

Several papers do consider some form of ground truth. Six
papers use it as their main performance measure, marked 'G' in the
table. By virtue of their choice of benchmark programs, they are able to 
map crashing inputs to their root cause perfectly. 
Eight other papers, marked 'G*' in the table, make some
effort to triage crashes to identify their root cause, but do so
imperfectly.  Typically, such triage is done as a `case study' and is
often neither well founded nor complete---ground truth is not used as
the overall (numeric) performance measure.

In the next three subsections we discuss performance measures
in detail, showing why using heuristics rather than actual ground truth to
compare fuzzer performance can lead to misleading or wrong
conclusions. In lieu of measuring bugs directly, nearly half of the
papers we examined consider a fuzzer's ability to execute
(``cover'') significant parts of a target program. This measurement is
potentially more generalizable than bug counts, but is not a
substitute for it; we discuss it at the end of the section.

\subsection{Ground Truth: Bugs Found}

The ultimate measure of a fuzzer is the number of distinct bugs that
it finds. If fuzzer $A$ generally finds more bugs than baseline $B$
then we can view it as more effective. A key question is: What is a
(distinct) bug? This is a subjective question with no
easy answer. 

We imagine that a developer will ultimately use
a crashing input to debug and fix the target program so that the crash
no longer occurs. That fix will probably not be specific to the input,
but will generalize. For example, a bugfix might consist of a
length check to stop a buffer overrun from occurring---this will work
for all inputs that are too long. As a result, if target $p$
crashes when given input $I$, but no longer crashes when the bugfix is
applied, then we can associate $I$ with the bug addressed by the
fix~\cite{Chen:2013:TCF:2491956.2462173}. Moreover, if inputs $I_1$
and $I_2$ both induce a crash on $p$, 
but both no longer do so once the bugfix is applied, we know that both
\emph{identify the same bug} (assuming the fix is suitably
``minimal''~\cite{whatbug}). 

When running on target programs with known bugs, we have direct access
to ground truth. Such programs might be older versions with bugs that
have since been fixed, or they might be synthetic programs or programs
with synthetically introduced bugs. Considering the former category,
we are aware of no prior work that uses old programs and their
corresponding fixes to completely triage crashes according to ground
truth. In the latter category, nine papers use synthetic suites in
order to determine ground truth. The most popular suites are CGC
(\emph{Cyber Grand Challenge})~\cite{cgc} and
LAVA-M~\cite{DBLP:conf/sp/Dolan-GavittHKL16}; we discuss these more in
the next section. For both, bugs have been injected into the original
programs in a way that triggering a particular bug produces a telltale
sign (like a particular error message) before the program crashes. As
such, it is immediately apparent which bug is triggered by the
fuzzer's generated input. If that bug was triggered before, the input
can be discarded. Two other papers used hand-selected programs
with manually injected vulnerabilities.

\subsection{AFL Coverage Profile}
\label{sec:afl-profile}

When ground truth is not available, researchers commonly employ 
heuristic methods de-duplicate crashing inputs.
The approach taken by AFL, and used by 7 papers
in Table~\ref{table:sum} (marked 'C'), is to consider inputs that have
the same code \emph{coverage profile} as equivalent. AFL will consider 
a crash ``unique'' if the edge coverage for that crash either contains
an edge not seen in any previous crash, or, is missing an edge that is
otherwise in all previously observed crashes.\footnote{AFL also
  provides a utility, \texttt{afl-cmin}, which can be 
  run offline to ``prune'' a corpus of inputs into a minimal corpus. Specifically,
the \prog{afl-cmin} algorithm keeps inputs that contain edges not contained
by any other inputs trace. This is different than the AFL on-line
algorithm, which also retains inputs missing edges that other inputs'
traces have. Only one prior paper that we know of, Angora~\cite{angora}, ran
\texttt{afl-cmin} on the final set of inputs produced by AFL; the rest
relied only on the on-line algorithm, as we do.}

\lstset{language=C,
                basicstyle=\small\ttfamily,
                keywordstyle=\color{blue}\ttfamily,
                stringstyle=\color{red}\ttfamily,
                commentstyle=\color{green}\ttfamily,
                morecomment=[l][\color{magenta}]{\#}
}

\begin{figure}[t]
\begin{center}
\begin{minipage}{2.5in}
\begin{lstlisting}
int main(int argc, char* argv[]) {
  if (argc >= 2) {
    char b = argv[1][0];
    if (b == 'a') crash();
    else          crash();
  }
  return 0;
}
\end{lstlisting}
\end{minipage}
\end{center}
\caption{How coverage-based deduplication can overcount}
\label{fig:ex-covprof}
\end{figure}

Classifying duplicate inputs based on coverage profile makes sense: it
seems plausible that two different bugs would have different coverage 
representations. 
On the other hand, it
is easy to imagine a single bug that can be triggered by runs with different
coverage profiles. For example, suppose the function \texttt{crash} in
the program in Figure~\ref{fig:ex-covprof} will segfault unconditionally. Though there
is but a single bug in the program, two classes of input will be
treated as distinct: those starting with an \lstinline{'a'} and those
that do not.

\paragraph*{Assessing against ground truth}
How often does this happen in practice? We examined the crashing inputs
our fuzzing runs generated for \prog{cxxfilt} using AFL and AFLFast. 
Years of development activity have occurred on this code
since the version we fuzzed was released, so (most of) the bugs that our fuzzing
found have been patched. We used \texttt{git} to identify commits that
change source files used to compile \prog{cxxfilt}. Then, we built every
version of \prog{cxxfilt} for each of those commits. This produced 
812 different versions of \prog{cxxfilt}. Then, we ran every crashing input
(57,142 of them) on each different version of \prog{cxxfilt}, recording 
whether or not that version crashed. If not, we consider the
input to have been a manifestation of a bug fixed by that program
version. 

To help ensure that our triaging results are trustworthy, we took two
additional steps. First, we ensured that non-crashing behavior was not
incidental. Memory errors and other bug categories uncovered by
fuzzing may not always cause a crash when triggered. For
example, an out-of-bounds array read will only crash the program if
unmapped memory is accessed. Thus it is possible that a commit could
change some aspect of the program that eliminates a crash without
actually fixing the bug. To address this issue, we compiled each
\prog{cxxfilt} version with Address Sanitizer and Undefined Behavior Sanitizer
(ASAN and UBSAN)~\cite{serebryany2012addresssanitizer}, which adds dynamic
checks for various errors including memory errors. We considered the
presence of an ASAN/UBSAN error report as a ``crash.''

Second, we ensured that each bug-fixing commit corresponds to a single
bugfix, rather than several. To do so, we manually inspected every
commit that converted a crashing input to a non-crashing one, judging
whether we believed multiple distinct bugs were being fixed (based on
principles we developed previously~\cite{whatbug}). If so, we manually split
the commit into smaller ones, one per fix. In our experiments, we only
had to do this once, to a commit that imported a batch of changes from
the \prog{libiberty} fork of \prog{cxxfilt} into the main
trunk.\footnote{\url{https://github.com/gcc-mirror/gcc/tree/master/libiberty}}
We looked at the individual \prog{libiberty}
commits that made up this batch to help us determine how to split it
up. Ultimately we broke it into five distinct bug-fixing
commits. 

\begin{figure*}[t]
 \centering
\includegraphics[height=8cm,width=.8\linewidth]{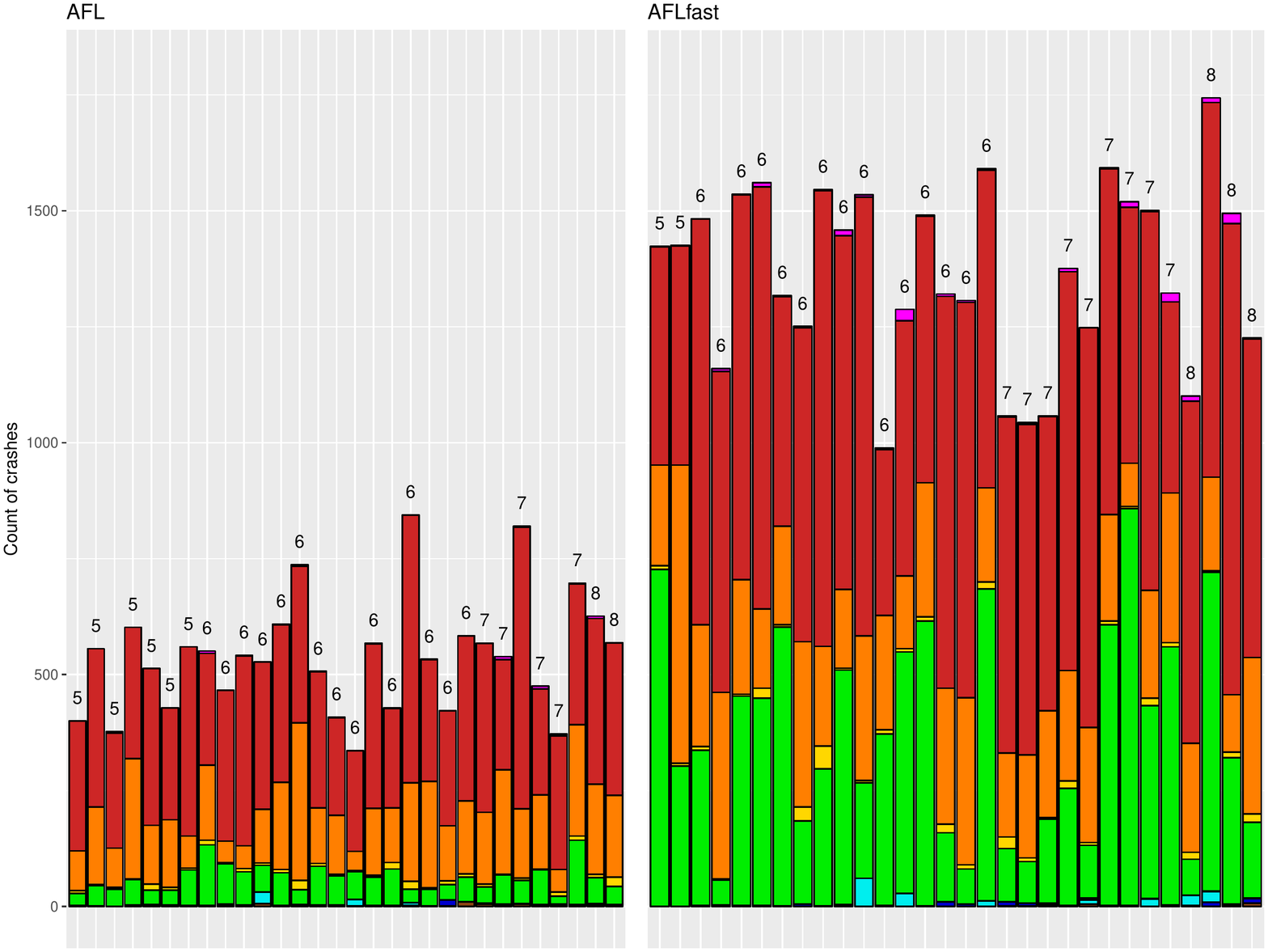}
\caption{Crashes with unique bugs found per run for \prog{cxxfilt}.
         Each bar represents an independent run of either AFL or AFLfast.
         The height of the bar is the count of crashing inputs discovered during 
         that run. Each bar is divided by color, clustering inputs with other inputs that 
         share the same root cause. 
         Number of unique bugs is indicated above each bar.}
\label{fig:dedup-cxxfilt}
\end{figure*}

Our final methodology produced 9 distinct bug-fixing commits, leaving a
small number of inputs that still crash the current version of
\prog{cxxfilt}. Figure~\ref{fig:dedup-cxxfilt} organizes these
results. 
Each bar in the graph represents a 24-hour fuzzing trial carried out by
either AFL or AFLFast.\footnote{We show each trial's data
  individually, rather than collecting it all together, because AFL's
  coverage-based metric was applied to each trial run, not all runs together.}
For each of these, the magnitude of the bar on the y axis is the total number of 
``unique'' (according to coverage profile) crash-inducing inputs, while the bar is segmented by which of these 
inputs is grouped with a bug fix discovered by our ground truth analysis. Above
each bar is the total number of bugs discovered by that run (which is
the number of compartments in each bar). 
The runs are ordered by the number of unique bugs found in the run. 

We can see that there is at best
a weak correlation between the number of bugs found during a run and the 
number of crashing inputs found in a run. Such a correlation would
imply a stronger upward trend of crash counts when moving left to
right. We can also see that AFLFast generally found many more
``unique'' crashing inputs than AFL but the number of
bugs found per run is only slightly higher. Mann Whitney
finds that the difference in crashes is statistically significant,
with a $p$-value of $10^{-10}$, but the difference in \emph{bugs} is
not (but is close)---the $p$-value is $0.066$. 


\paragraph*{Discussion}
Despite the steps we took to ensure our triage matches ground truth,
we may still have inflated or reduced actual bug counts. As
an example of the former, we note that ASAN/UBSAN is not guaranteed to catch
all memory safety violations, so we may have attributed an
incidental change to a bugfix. We found a few cases where we couldn't
explain why a commit fixed a crash, and so did not associate the commit with
a bug. On the other hand, we might have failed
to differentiate multiple bugfixes in a single commit, either by
mistake or in the eyes of an observer whose judgment differs
from our own. In any case, the magnitude of the difference
between our counts and ``unique crashes'' means that the top-level
result---that ``unique crashes'' massively overcount the number of true
bugs---would hold even if the counts changed a little.

Had we used ground truth measure in all of our experiments, it might
have changed the character of the 
results in Sections~\ref{sec:sta-sig}--\ref{sec:timeout}.
For example, the performance variations within
a configuration due to randomness (e.g., Figure~\ref{fig:result}) may
not be as stark when counting bugs rather than ``unique'' crashing
inputs. In this case, our advice of carrying out multiple trials is
even more important, as small performance differences between fuzzers
$A$ and $B$ may require many trials to discern. It
may be that performance differences due to varying a seed
(Figure~\ref{fig:ffmpeg}) may also not be as stark---this would be true if
one seed found hundreds of crashes and another found far fewer, but
in the end all crashes corresponded to the same bug. There may also be
less performance variation over time when bugs, rather than crashes,
are counted (Figure~\ref{fig:timeout}). On the other hand, it is also
possible that we would 
find \emph{more} variation over time, and/or with different seeds, 
rather than less. In either case, we believe our results in
Sections~\ref{sec:sta-sig}--\ref{sec:timeout} raise sufficient
concern that our advice to test with longer timeouts and a variety of
seeds (including the empty seed) should be followed unless and until
experimental results with ground truth data shed more light on the
situation. 




\subsection{Stack hashes}

Another common, heuristic de-duplication technique is stack
hashing~\cite{Molnar:2009:DTG:1855768.1855773}. Seven papers we
considered use this technique (marked 'S' in Table~\ref{table:sum}). 
The idea is the following.  Suppose that our buffer overrun bug is in
a function deep inside a larger program, e.g., in a library
routine. Assuming that the overrun induces a segfault immediately, the
crash will always occur at the same place in the program. More
generally, the crash might depend on some additional program context;
e.g., the overrun buffer might only be undersized when it is created
in a particular calling function. In that case, we might look at the
\emph{call stack}, and not just the program counter, to map a crash to
a particular bug. To ignore spurious variation, we focus on return
addresses normalized to their source code location. 
Since the part of the stack closest to the top is
perhaps the most relevant, we might only associate the most recent $N$
stack frames with the triggering of a particular bug. 
($N$ is often chosen to be between 3 and 5.)
These frames could be hashed for quick comparison to prior bugs---a
\emph{stack hash.} 

Stack hashing will work as long as relevant context is unique, and
still on-stack at the time of crash. But it is easy to see situations
where this does not hold---stack hashing can end up both undercounting
or overcounting true bugs. Consider the code in Figure~\ref{fig:ex-stackhash}, which
has a bug in the \texttt{format} function that corrupts a string
\texttt{s}, which ultimately causes the \texttt{output} function to crash
(when \texttt{s} is passed to it, innocently, by the \texttt{prepare}
function). The \texttt{format} function is called separately by
functions \texttt{f} and \texttt{g}.

\begin{figure}[t]
\begin{center}
\begin{minipage}{2.5in}
\begin{lstlisting} 
void f() { ... format(s1); ... }
void g() { ... format(s2); ... }
void format(char *s) {
  //bug: corrupt s
  prepare(s);
}
void prepare(char *s) {
  output(s);
}
void output(char *s) {
  //failure manifests
}
\end{lstlisting}
\end{minipage}
\end{center}
\caption{How stack hashing can over- and undercount bugs}
\label{fig:ex-stackhash}
\end{figure}

Suppose we fuzz this program and generate inputs that induce two
crashes, one starting with the call from \texttt{f} and the other
starting with the call from \texttt{g}. Setting $N$ to the top 3 frames,
the stack hash will correctly recognize that these two inputs
correspond to the same bug, since only \texttt{format}, \texttt{prepare}
and \texttt{output} will be on the stack. Setting $N$ to 5, however,
would treat the inputs as distinct crashes, since now one stack
contains \texttt{f} and the other contains \texttt{g}. On the other hand,
suppose this program had another buggy function that also corrupts
\texttt{s} prior to passing it to \texttt{prepare}. Setting $N$ to 2 would
improperly conflate crashes due to that bug and ones due the buggy
\texttt{format}, since only the last two functions on the stack would be
considered. 

\paragraph*{Assessing against ground truth}
We measured the effectiveness of stack hashing by comparing its
determinations against the labels for
bugs that we identified in the prior experiment. Our implementation of
stack hashing uses Address Sanitizer to produce a stack trace for each
crashing input 
to \prog{cxxfilt}, and chooses $N$ as the top 3 entries on the stack for
hashing. 

\begin{table}[t]
\centering
  \caption{Stack hashing results for \prog{cxxfilt}. The first column specifies the 
           label we assign based testing progressive versions of \prog{cxxfilt}. 
           The second column specifies the number of distinct stack hashes among the
           inputs assigned to the ground truth label. The third column counts how many
           of the stack hashes from the second column appear only with those inputs 
           grouped by the label in the first column, while the fourth column counts
           how many stack hashes appear in other labels. The final column counts
           the number of distinct inputs in a label.}
\label{stack-hashing}
\label{tab:hashing}
\begin{tabular}{@{}lllll@{}}
\toprule
  Bug & \# Hashes & Matches & False Matches & Input count  \\ \midrule
  A   & 9               & 2       & 7             & 228          \\
  B   & 362             & 343     & 19            & 31,103        \\
  C   & 24              & 21      & 3             & 106          \\
  D   & 159             & 119     & 40            & 12,672        \\
  E   & 15              & 4       & 11            & 12,118        \\
  F   & 15              & 1       & 14            & 232          \\
  G   & 2               & 0       & 2             & 2            \\
  H   & 1               & 1       & 0             & 568          \\
  I   & 4               & 4       & 0             & 10           \\
  unfixed              & 28              & 12      & 16            & 98           \\ 
  unknown              & 4               & 0       & 4             & 4            \\ \bottomrule
\end{tabular}
\end{table}

Our analysis discovered that stack hashing is far more effective at
deduplicating inputs than coverage profiles, but would still
over-count the number of bugs discovered. Table~\ref{tab:hashing}
shows the results of the comparison of stack hashing to the labels we
identified. As an example, consider label B, which represents 31,103
inputs (column 5). Of those inputs, 362 distinct stack hashes were
produced (column 2). If the stack hash metric was the only knowledge
we had about the distribution of bugs in \prog{cxxfilt}, we would
claim to have discovered two orders of magnitude more bugs than we
actually did. On the other hand, stack hashing seems to do very well
for label H: one hash matched all 568 inputs. In sum, across all runs, 595 hashes
corresponded to 9 bugs, an inflation of $66\times$, as compared to
57,044 coverage profile-unique inputs for 9 bugs, an inflation of
$6339\times$.\footnote{The table tabulates crashing inputs across all
  trials put together: if instead you consider the stack hashes taken
  on a per-run basis (as in Figure~\ref{fig:dedup-cxxfilt}), the
  results will be somewhat different, but the overall trends should
  remain the same.}
4 crashing inputs were each associated with their own ``fixing'' commit, 
but when we inspected the respective code changes we could not see 
why the changes should fix a crash. As such,
we have listed these inputs in Table~\ref{tab:hashing} as ``unknown.''
ASAN/UBSAN does not detect all possible undefined behaviors, so
it may be that a code or data layout 
change between compilations or some other form of non-determinism 
is suppressing the crashing behavior. A compiler bug is also a possibility.
We are continuing to investigate.

While stack hashing does not overcount bugs nearly as much as AFL
coverage profiles, it has the serious problem that hashes are not
unique. For example, only 
343 of those for label B matched \emph{only} inputs associated with B
(column 3). The remaining 19 \emph{also} matched some other crashing
input (column 4). As such, these other inputs would be wrongly
discarded if stack hashing had been used for de-duplication. Indeed,
for label G, there is no unique hash (there is a 0 in column 3)---it only falsely matches. 
Overall, about 16\% of hashes were 
non-unique.\footnote{This value was computed by summing the total
  distinct number of hashes that show up in more than one row (a lower
  bound of the total in column 4) and dividing by the total of distinct hashes
  overall (a lower bound of the total in column 2).} As such, stack hashing-based
deduplication would have discarded these bugs. 

\paragraph*{Discussion}
Table~\ref{stack-hashing} shows another interesting trend also
evident, but less precisely, in Figure~\ref{fig:dedup-cxxfilt}. Some
bugs are triggered by a very small number of inputs, while others by a
very large number. Bugs G and I each
correspond to only 2 or 10 inputs, while bugs B, D, and E correspond to
more than 10K inputs. Prior fuzzing studies have found similar bug
distributions~\cite{Chen:2013:TCF:2491956.2462173}. While Table~\ref{stack-hashing} combines all
inputs from all trials, considering each trial individually (as per
Figure~\ref{fig:dedup-cxxfilt}) we find that no single run found all
9 bugs; all runs found bugs B, D, E, but no run found more than 5
additional bugs. 

An important open question is whether the trends we observe here with
\prog{cxxfilt} hold for other target programs. To answer this question would require
more ``ground truth'' analysis of the flavor we have carried out
here. Assuming they do hold, we draw two tentative conclusions. First,
the trends reinforce the problem with bug heuristics: in the presence of
``rare'' inputs, the difference between finding 100 crashing
inputs and 101 (an apparently insignificant difference) could
represent finding 1 or 2 unique bugs (a significant one). Second,
fuzzers might benefit from an algorithmic trick employed by SAT
solvers: randomly ``reboot'' the search
process~\cite{Ryvchin:2008:LR:1789854.1789879} by
discarding some of the current state and starting again with the
initial seed, thus 
simulating the effect of running separate trials.
The challenge would be to figure out what fuzzer state to retain across
reboots so as to retain important knowledge but avoid
getting stuck in a local minimum. 

\paragraph*{Related Work}
Recent work by \citet{vantonder18sbc} also experimentally assesses the
efficacy of stack hashing and coverage profiles against ground
truth. Like us, they defined ground truth as single conceptual bugs
corrected by a particular code patch. They compared how well coverage
profiles and stack hashes approximate this ground truth. Like us, they
found that both tended to overcount the number of true bugs. As they
consider different patches and target programs, their study is
complementary to ours. However, their set of crashing inputs was
generated via mutations to an initial known crashing input, rather
than via a normal fuzzing process. As such, their numbers do not
characterize the impact of poor deduplication strategies in typical
fuzzing use-cases, as ours do. 

\citet{pham17bucketing} also studied how stack hashes, for $N=1$ and
$N=\infty$, can over- and under-count bugs identified through
symbolic execution. Their interest was a comparison against their own
de-duplication technique, and so their study did not comprehensively
consider ground truth.

\subsection{Code Coverage}

Fuzzers are run to find bugs in programs. A fuzzer that runs for a long 
period of time and finds no bugs would be seen as unsuccessful by its 
user. It seems logical to evaluate a fuzzer based on the number of bugs
that fuzzer finds. However, just because a fuzzer does not find a bug
may not tell us the whole story about the fuzzer's efficacy. Perhaps
its algorithm is sound but there are few or no bugs to find, and the
fuzzer has merely gotten unlucky. 

One solution is to instead (or also) measure the improvement in
code coverage made by fuzzer $A$ over baseline $B$. Greybox fuzzers
already aim to optimize coverage as part of the \texttt{isInteresting}
function, so surely showing an improved code coverage would indicate
an improvement in fuzzing. This makes sense. To find a
crash at a particular point in the program, that point in the program
would need to execute. Prior studies of test suite effectiveness also 
suggest that higher coverage correlates with bug finding
effectiveness~\cite{7081877,Gopinath:2014:CCS:2568225.2568278}. Nearly
half of the papers we considered measured code coverage; FairFuzz 
\emph{only} evaluated performance using code (branch)
coverage~\cite{lemieux2017fairfuzz}.  

However, there is no \emph{fundamental} reason that maximizing code coverage
is directly connected to finding bugs. While the general efficacy of
coverage-guided fuzzers over black box ones implies that there's a
strong correlation, particular algorithms may eschew higher coverage
to focus on other signs that a bug may be present. For example,
AFLGo~\cite{bohme2017directed} does not aim to increase coverage
globally, but rather aims to focus on particular, possibly error-prone
points in the program. Even if we assume that coverage and bug finding
are correlated, that correlation may be
weak~\cite{Inozemtseva:2014:CSC:2568225.2568271}. As such, a substantial
improvement in coverage may yield merely a negligible improvement in bug
finding effectiveness.

In short, we believe that code coverage makes sense as a
secondary measure, but that ground truth, according to bugs
discovered, should always be primary.

\section{Target Programs}
\label{sec:bm}

We would like to establish that one fuzzing algorithm is
\emph{generally} better than another, i.e., in its ability to find
bugs in any target program drawn from a (large) population.
Claims of generality are usually made by testing the
fuzzer on a benchmark suite that purports to represent the
population. The idea is that good performance on the suite should
translate to good performance on the population. How should we choose
such a benchmark suite? 


Recent published works have considered a wide variety of benchmark
programs. Broadly, these fall into two categories, as shown in the
second column in Table \ref{table:sum}: real programs and 
artificial programs (or bugs). Examples of the former include the
Google fuzzer test suite (``G'') \cite{fuzzer-test-suite} and ad hoc
selections of real programs (``R''). The latter comprises
CGC (``C'') \cite{cgc}, LAVA-M (``L'')
\cite{DBLP:conf/sp/Dolan-GavittHKL16}, and hand-selected programs with
synthetically injected bugs (``S''). Some papers' benchmarks drew from
both categories (e.g., VUzzer \cite{rawat2017vuzzer} and Steelix
\cite{li2017steelix}). As we discuss below, no existing
benchmark choice is entirely satisfying, thus leaving open the
important question of developing a good fuzzing benchmark.

\subsection{Real programs} 

According to Table \ref{table:sum}, nearly all papers used some
real-world programs in their evaluations. 
Two of these papers~\cite{Xu:2017:DNO:3133956.3134046,
  DBLP:conf/raid/ShastryLFTYRSSF17} used the Google Fuzzer Test
suite~\cite{fuzzer-test-suite}, a set of real-world programs and
libraries coupled with harnesses to focus fuzzing on a set of known
bugs. The others evaluated on a hand selected set of real-world
programs. 

We see two problems with the way that real programs have been used as
fuzzing targets. First, most papers consider only a small number of
target programs without clear justification of their
representativeness. The median number of programs, per Table
\ref{table:sum}, is seven. Sometimes a small count is justified; e.g.,
IMF was designed specifically to fuzz OS kernels, so its evaluation on
a single ``program,'' the MacOS kernel, is still interesting. On the
other hand, most fuzzers aim to apply to a larger population (e.g., all
file processing programs), so 7 would seem to be a small number. A
positive outlier was FuzzSim, which used a large set of programs (more
than 100) and explained the methodology for collecting them.

As evidence of the threat posed by a small number of insufficiently
general targets, consider the experimental results reported in
Figure~\ref{fig:result}, which match the results of B\"{o}hme et 
al~\cite{Bohme:2016:CGF:2976749.2978428}. 
The first row of the figure shows results for \prog{nm},
\prog{objdump} and \prog{cxxfilt}, which were the three programs in
which B\"{o}hme et al found crashes.\footnote{Figure 6 of their paper
  presents a similar series of plots. The differences in their plots
  and ours are the following: they plot the results on log scale for
  the Y axis; they consider six-hour trials rather than 24-hour
  trials; and they do not plot median and confidence intervals
  computed over 30+ runs, but rather plot the mean of 8 runs. They
  also use different versions of AFL and AFLFast.}
Focusing our attention on these programs suggests that AFLFast is
uniformly superior to AFL in crash finding ability.
However, if we look at the second row of the figure, the story is not
as clear. For both \prog{FFmpeg} and \prog{gif2png}, two
programs used in other fuzzing evaluations, the Mann Whitney U
test shows no statistical difference between AFL and AFLFast.
Including these programs in our assessment
weakens any claim that AFLFast is an improvement over AFL.

The second problem we see with the use of real programs to date is
that few papers use the same targets, at the same versions. As such, it is hard
to make even informal comparisons across different papers. One overlapping
set of targets were \prog{binutils} programs, used in several
evaluations~\cite{Bohme:2016:CGF:2976749.2978428, lemieux2017fairfuzz,
  bohme2017directed, angora}. Multiple papers also considered \prog{FFmpeg}
and \prog{gif2png}~\cite{Woo:2013:SBM:2508859.2516736,
  rawat2017vuzzer, Cha:2015:PMF:2867539.2867674,
  Rebert:2014:OSS:2671225.2671280, DBLP:conf/cis/ZhangYFT17}. However,
none used the same versions.
For example, the versions of \prog{binutils} were different in these papers:
AFLFast \cite{Bohme:2016:CGF:2976749.2978428} and AFLGo \cite{bohme2017directed} used 2.26;
FairFuzz \cite{lemieux2017fairfuzz} used 2.28; Angora \cite{angora} used 2.29.

The use of Google Fuzzer Suite would seem to address both issues: it
comprises 25 programs with known bugs, and is defined independently of
any given fuzzer. On the other hand, it was designed as a kind of
regression suite, not necessarily representative of fuzzing ``in the
wild;'' the provided harnesses and seeds mostly intend that fuzzers
should find the targeted bugs within a few seconds to a few minutes.

\subsection{Suites of artificial programs (or bugs)} 

Real programs are
fickle in that the likelihood that bugs are present depends on many
factors. For example, programs under active development may well have
more bugs than those that are relatively stable (just responding to
bug reports). In a sense, we do not care about any particular set of
programs, but rather a representative set of programming
(anti)patterns in which bugs are likely to crop up. Such patterns
could be injected artificially. There are two popular suites that do
this: CGC, and LAVA-M.

The CGC suite comprises 296 buggy programs produced as part of DARPA's
Cyber Grand Challenge \cite{cgc}. This suite was specifically designed to
evaluate bug finding tools like fuzz testers---the suite's programs
perform realistic functions and are seeded with exploitable bugs. LAVA
(which stands for Large-scale Automated Vulnerability Addition) is a
tool for injecting bugs into known programs \cite{DBLP:conf/sp/Dolan-GavittHKL16}. The tool is designed to
add crashing, input-determinate bugs along feasible paths. The LAVA
authors used the tool to create the LAVA-M suite, which comprises four
bug-injected \prog{coreutils} programs: \prog{base64}, \prog{md5sum},
\prog{uniq}, and \prog{who}. Unlike the CGC programs, which have very
few injected bugs, the LAVA-M programs have many: on the order of a
few dozen each for the first three, and more than 2000 for
\prog{who}. For both suites, if a fuzzer triggers a bug, there is a 
telltale sign indicating which one it is, which is very useful for
understanding how many bugs are found from the total possible.

CGC and LAVA-M have gained popularity as the benchmark choices
for evaluating fuzzers since their introduction. Within the past
two years, CGC and LAVA-M have been used for evaluating 4 and 5 fuzzers,
respectively. VUzzer \cite{rawat2017vuzzer}, Steelix \cite{li2017steelix}, and T-Fuzz \cite{tfuzz}
used both benchmarks in their evaluation. However, sometimes the CGC
benchmark was subset: Driller \cite{stephens2016driller}, VUzzer \cite{rawat2017vuzzer},
and Steelix \cite{li2017steelix} were evaluated on 126, 63, and 17 out of
the 296 programs, respectively.

While CGC programs are hand-designed to simulate reality, this simulation
may be imperfect: Performing well on the CGC programs may fail to
generalize to actual programs. For example, the average size of the
CGC cqe-challenge programs was (only) 1774 lines of code, and many
programs use telnet-style, text-based protocols. Likewise, LAVA-M injected bugs may not
sufficiently resemble those found ``in the wild.'' The incentives and circumstances behind
real-world software development may fail to translate to synthetic
benchmarks which were specifically designed to be insecure. The LAVA
authors write that, ``A significant chunk of future work for LAVA
involves making the generated corpora look more like the bugs that are
found in real programs.''  Indeed, in recent experiments~\cite{lava-m-blog},
they also have shown that relatively simple techniques can effectively
find all of the LAVA-M bugs, which follow a simple pattern. We are
aware of no study that independently 
assesses the extent to which these suites can be considered ``real''
or ``general.''

\subsection{Toward a Fuzzing Benchmark Suite}

Our assessment leads us to believe that there is a real need for a
solid, independently defined benchmark suite, e.g., a DaCapo
\cite{Blackburn:2006:DBJ:1167473.1167488} or
SPEC\footnote{\url{https://www.spec.org/benchmarks.html}} for fuzz
testing. This is a big enough task that we do not presume to take it
on in this paper. It should be a community effort. That said, we do
have some ideas about what the result of that effort might look like.

First, we believe the suite should have a selection of programs with
clear indicators of when particular bugs are found, either because
bugs are synthetically introduced (as in LAVA-M and CGC) or because
they were previously discovered in older versions (as in our ground
truth assessment in Section~\ref{sec:afl-profile}). Clear knowledge of
ground truth avoids overcounting inputs that correspond to the same
bug, and allows for assessing a tool's false positives and false
negatives. We lean toward using real programs with known bugs simply
because their ecological validity is more assured. 

Second, the suite should be large enough (both in number of programs,
and those programs' sizes) to represent the overall target
population. How many programs is the right number? This is an open
question. CGC comprises $\sim{300}$ small programs; Google Fuzzer
Suite has 25; most papers used around 7. Our feeling is that 7 is too
small, but it might depend on which 7 are chosen. Perhaps 25 is closer
to the right number.

Finally, the testing methodology should build in some defense against
overfitting. If a static benchmark suite comes into common use, tools
may start to employ heuristics and strategies that are not of
fundamental advantage, but apply disproportionately to the benchmark
programs. One way to deal with this problem is to have a fixed
standard suite and an ``evolvable'' part that changes relatively
frequently. One way to support the latter is to set up a fuzzing
competition, similar to long-running series of SAT solving
competitions.\footnote{\url{http://www.satcompetition.org/}} One
effort in this direction is Rode0day, a recurring bug finding
competition.\footnote{\url{https://rode0day.mit.edu/}} Since the
target programs would not be known to fuzzing researchers in advance,
they should be incentivized to develop general, reusable
techniques. Each competition's suite could be rolled into the static
benchmark, at least in part, to make the suite even more robust. One
challenge is to regularly develop new targets that are ecologically valid. For
example, Rode0day uses automated bug insertion techniques to
which a tool could overfit (the issue discussed above for LAVA).

\section{Conclusions and Future Work}
\label{sec:conc}

Fuzz testing is a promising technology that has been used to uncover
many important bugs and security vulnerabilities. This promise has
prompted a growing number of researchers to develop new fuzz testing
algorithms. The evidence that such algorithms work is primarily
experimental, so it is important that it comes from a well-founded experimental
methodology. In particular, a researcher should run their algorithm
$A$ on a general set of target programs, using a meaningful set of
configuration parameters, including the set of input seeds and
duration (timeout), and compare against the performance of a baseline
algorithm $B$ run under the same conditions, where performance is
defined as the number of (distinct) bugs found. $A$ and $B$ must be
run enough times that the inherent randomness of fuzzing is accounted
for and performance can be judged via a statistical test.

In this paper, we surveyed 32 recent papers and analyzed their
experimental methodologies. We found that no paper completely follows
the methodology we have outlined above. Moreover, results of
experiments we carried out using
AFLFast~\cite{Bohme:2016:CGF:2976749.2978428} (as $A$) and
AFL~\cite{afl} (as $B$) illustrate why not following this methodology
can lead to misleading or weakened conclusions. We found that 
\begin{itemize}[leftmargin=.2in]
\item Most papers failed to perform multiple runs, and those that
  did failed to account for varying performance by using a statistical
  test. This is a problem because our experiments showed that
  run-to-run performance can vary substantially.  
\item Many papers measured fuzzer performance not by counting distinct
  bugs, but instead by counting ``unique crashes'' using
  heuristics such as AFL's coverage measure and stack hashes. This is
  a problem because experiments we carried out showed that the
  heuristics can dramatically over-count the number of bugs, and indeed
  may suppress bugs by wrongly grouping crashing inputs. This means
  that apparent improvements may be modest or illusory. Many papers made
  some consideration of root causes, but often as a ``case study''
  rather than a performance assessment.
\item Many papers used short timeouts, without justification. Our
  experiments showed that longer timeouts may be needed to paint a
  complete picture of an algorithm's performance.
\item Many papers did not carefully consider the impact of seed
  choices on algorithmic improvements. Our experiments showed that
  performance can vary substantially depending on what seeds are
  used. In particular, two different non-empty inputs need not
  produce similar performance, and the empty seed can work better than
  one might expect.
\item Papers varied widely on their choice of target programs. A
  growing number are using synthetic suites CGC and/or LAVA-M, which
  have the advantage that they are defined independently of a
  given algorithm, and bugs found by fuzzing them can be reliably counted
  (no crash de-duplication strategy is needed). Other papers often
  picked small, disjoint sets of programs, making it difficult to
  compare results across papers. Our experiments showed AFLFast performs well on the
  targets it was originally assessed against, but performed no better
  than AFL on two targets used by other papers.
\end{itemize}

\noindent
Ultimately, our experiments roughly matched the positive results of the
original AFLFast paper~\cite{Bohme:2016:CGF:2976749.2978428}, but by
expanding the scope of the evaluation to different seeds, longer
timeouts, and different target programs, evidence of AFLFast's 
superiority, at least for the versions we tested, was weakened. The fact that
heuristic crash de-duplication strategies are of questionable value
further weakens our confidence in claims of improvement. We
believe the same could be said of many prior papers---all suffer from
problems in their evaluation to some degree. As such, a key conclusion
of this paper is that the fuzzing community needs to start carrying
out more rigorous experiments in order to draw more reliable
conclusions. 


Specifically, we recommend that fuzz testing evaluations should have
the following elements:
\begin{itemize}[leftmargin=.2in]

\item multiple trials with statistical tests to distinguish distributions;

\item a range of benchmark target programs with known bugs (e.g.,
  LAVA-M, CGC, or old programs with bug fixes);

\item measurement of performance in terms of known bugs, rather than
  heuristics based on AFL coverage profiles or stack hashes; block or
  edge coverage can be used as a secondary measure;

\item a consideration of various (well documented) seed choices including empty seed;

\item timeouts of at least 24 hours, or else justification for less,
  with performance plotted over time.
\end{itemize}

We see (at least) three important lines of future work. First, we believe there is
a pressing need for well-designed, well-assessed benchmark suite, as
described at the end of the last section.
Second, and related, it would be worthwhile to carry out a larger
study of the value of crash de-duplication methods on the results of
realistic fuzzing runs, and potentially
develop new methods that work better, for assisting with triage and
assessing fuzzing when ground truth is not known. Recent work
shows promise~\cite{pham17bucketing,vantonder18sbc}.
Finally, it would be interesting to explore enhancements to
the fuzzing algorithm inspired by the observation that no single
fuzzing run found all true bugs in \prog{cxxfilt}; ideas from other
search algorithms, like SAT solving
``reboots''~\cite{Ryvchin:2008:LR:1789854.1789879}, might be brought
to bear. 

\paragraph{Acknowledgments}
We thank Marcel B\"{o}hme and Abhik Roychoudhury for their help with
AFLFast. We thank the anonymous reviewers, Michelle Mazurek, Cornelius
Aschermann, Luis Pina, Jeff Foster, Ian Sweet, the participants of the
ISSISP'18 summer school, and our shepherd Mathias Payer for helpful
comments and suggestions on drafts of this work. This research was
supported in part by the National Science Foundation grants
CNS-1563722 and CNS-1314857, and DARPA under contracts
FA8750-15-2-0104 and FA8750-16-C-0022, and a Google Research Award.

\balance

\bibliographystyle{ACM-Reference-Format}
\bibliography{main}

\end{document}